\documentclass[a4paper,man,natbib]{apa6}

\usepackage[english]{babel}
\usepackage[utf8x]{inputenc}
\usepackage{amsmath}
\usepackage{graphicx}
\usepackage[colorinlistoftodos]{todonotes}
\usepackage{hyperref} 
\usepackage{color}    
\usepackage{multirow} 
\usepackage{placeins} 

\newcommand{\norm}[1]{\left\lVert#1\right\rVert} 

\newcommand{\pathFIG}{./figures}

\title{High-dimensional Imputation for the Social Sciences: a Comparison of State-of-the-art Methods}
\shorttitle{High-dimensional imputation}
\fourauthors
	{Edoardo Costantini}
	{Kyle M. Lang}
	{Tim Reeskens}
	{Klaas Sijtsma}
\fouraffiliations
	{Tilburg University, Department of Methodology and Statistics\\Contact the author: e.costantini@tilburguniversity.edu}
	{Utrecht University, Department of Methodology and Statistics}
	{Tilburg University, Department of Sociology}
	{Tilburg University, Department of Methodology and Statistics}
\keywords{
    Multiple Imputation,
    High-dimensionality,
    Regularized Regression,
    Principal components,
    CART,
    random forest
}

\abstract{
    Including a large number of predictors in the imputation model underlying a multiple imputation (MI)
procedure is one of the most challenging tasks imputers face.
A variety of high-dimensional MI techniques can help, but there has been
limited research on their relative performance.
In this study, we investigated a wide range of extant high-dimensional MI techniques
that can handle a large number of predictors in the imputation models and general missing data patterns.
We assessed the relative performance of seven high-dimensional MI methods with a Monte Carlo simulation study and a
resampling study based on real survey data.
The performance of the methods was defined by the degree to which they facilitate unbiased and confidence-valid
estimates of the parameters of complete data analysis models.
We found that using lasso penalty or forward selection to select the predictors used in the MI model and using principal
component analysis to reduce the dimensionality of auxiliary data produce the best results.

    }

\begin{document}
	\maketitle

    \setcounter{secnumdepth}{3} 

    \section{Introduction}

    Today’s social, behavioral, and medical scientists have access to large multidimensional data sets that can be
    used to investigate the complex roles that social, psychological, and biological factors play in
    shaping individual and societal outcomes.
    Large social scientific data sets---such as the World Values Survey and the European Values Study (EVS)---are easily
    accessible to researchers, but making use of the full potential of these data requires dealing with the crucial problem
    of multivariate missing data.

\subsection{The state of imputation in Sociology}

    Sociologists working with social surveys are usually interested in drawing inferential conclusions based on a substantively interesting analysis model.
    Generally, these analysis models require complete data, so the researcher must address any missing values before moving on to their substantive analysis.
    There are many possible missing data treatments from which to choose, and their relative strengths and weaknesses are covered elsewhere \citep[e.g.,][]{littleRubin:2002, enders:2010, vanBuuren:2018}.
    In this paper, we will focus on \cite{rubin:1987}'s multiple imputation (MI), which is one of the most effective ways of addressing missing values in survey data.

    MI is a three-step procedure that entails imputation, analysis, and pooling phases.
    The fundamental idea of the \emph{imputation phase} is to replace each missing data point with $d$ plausible values sampled from the posterior predictive distribution of the missing data, given the observed data.
    This phase generates $d$ completed versions of the original data set that are each analyzed separately during the \emph{analysis phase}, using any standard complete data analysis model.
    Finally, in the \emph{pooling phase}, the $d$ sets of estimates from the analysis models are pooled following Rubin's rules \citep{rubin:1987} to create a single set MI parameter estimates and standard errors.

    Missing values are one of the main factors impacting the quality of data gathered with surveys \citep{meyerEtAl:2015}, and nonresponse rates in large social survey have risen drastically over the last two decades \citep{brickWilliams:2013, masseyTourangeau:2013, williamsBrick:2018}.
    To explore how sociologists are addressing the issue of nonresponse in their research, we reviewed how missing data have been discussed in the articles published over the last five years in two leading sociological journals: American Journal of Sociology (AJS) and American Sociological Review (ASR).
    We found that of the 148 AJS research articles that mentioned using a survey, or some form of sample, for inferential analysis, 24 addressed the presence of missing values, and 17 conducted some form of imputation.
    Of these 17, only 13 performed MI, and among these 13, only three articles gave information on which predictors were used in the imputation models.
    Turning to ASR, the picture was similar.
    Of the 191 research articles published between January 2017 and January 2022 that met the inclusion criteria described above, 20 reported performing MI\@.
    Of these 20 articles, only six gave information regarding which predictors were used in the imputation models.
    Across the two journals, in the nine papers we found that described which predictors were used in the imputation models, the predominant choice was to use only the analysis model variables in the imputation model.

    In general, it seems that even when sociologists pay attention to the problem of missing values, little attention is given to which variables should be used in the imputation models.
    Similar conclusions were drawn in other literature reviews \citep{mustillo:2012, mustilloSoyoung:2015}.
    However, which variables to include in the imputation models is a crucial decision in MI\@.
    Leaving out important predictors of missingness can induce \emph{missing not at random} (MNAR) data \citep{collinsEtAl:2001}, while including good predictors can both correct for nonresponse bias and improve the efficiency of the parameters estimates \citep{collinsEtAl:2001, vonHippel:2013}.

\subsection{The challenge of specifying good imputation models}

    Specifying the imputation model is one of the most challenging steps in dealing with missing values.
    As described by \cite{vanBuurenEtAl:1999}, the task involves defining two aspects of the model: the model form (e.g., linear, logistic) and the predictor matrix (i.e., the set of predictors that enter the imputation model).
    The first choice is straightforward in virtually any imputation task, as it depends primarily on the measurement level of the variables under imputation.
    The second choice requires a careful selection process aimed at identifying the subset of variables that will be most useful in a given imputation model.

    Generally, the variables that will be part of the analysis model should also be included in the imputation model.
    When some analyzed variables (including transformations such as polynomials or interactions) are excluded from the imputation, the analysis and imputation models are said to be uncongenial \citep{meng:1994}.
    Such uncongeniality can lead to biased parameter estimates and invalid inferences.
    When designing an imputation model, the range of analysis models for which the resulting imputations will be congenial is an important consideration.
    In the methodological literature, this concept is known as the scope of the imputation model.
    \citet[p. 46]{vanBuuren:2018} distinguishes three typical imputation model scopes:
    \begin{itemize}

        \item \emph{Narrow Scope:} Narrowly scoped imputation models are matched to individual analyses.
        In such a scenario, the imputation is a customized pre-processing step intended to facilitate only a single analysis model.
        When imputing with a narrow scope, the primary objective is to ensure that all the variables in the analysis models (including relevant transformations) appear in the imputation model.
        An analyst who imputes their own data and plans to estimate only one model (or a single series of nested models) may wish to specify a narrow scoped imputation model.

        \item \emph{Intermediate Scope:} Imputation models with an intermediate scope are designed to support several different analysis models.
        The imputer will generally know approximately which analyses are intended but may not have an exhaustive list of all variables that will be analyzed.
        The objective is to design an imputation model that will be congenial with all planned and unplanned analysis models.
        Such analytic contexts frequently arise within research teams wherein several different analyses contribute to a larger research program.
        The evaluation of the Dating Matters intervention \citep{tharp:2012} is an example of one such research program.
        Due to the size and complexity of the data and the diversity of the intended analyses, treating the missing data in the Dating Matters evaluation took several months of dedicated work \citep{niolonEtAl:2019}.
        The resulting imputations were then used to support the substantive analyses by which various dimensions of the intervention were evaluated \citep[e.g.,][]{vivoloEtAl:2021, estefanEtAl:2021}.

        \item \emph{Broad Scope:} Imputation models with a broad scope are designed to create imputations that will be congenial to the most general set of analysis models feasible.
        The imputer cannot know beforehand which variables will be part of the analysis models, so the imputation models are designed to be general enough to accommodate a wide range of potential analyses.
        Practically speaking, the objective is to recreate the moments of the hypothetically fully observed data as closely as possible.
        \citet[p. 3]{rubin:1987} originally envisioned MI as a method using broadly scoped imputation models to treat publicly released data and argues that well-implemented MI can accommodate models that were not contemplated by the imputer \citep[p. 218]{littleRubin:2002}.
        Any data curation institution imputing data that are intended for public release will need imputation models with a broad scope.
        The Federal Reserve Board's Survey of Consumer Finances \citep{kennickell:1998} and the Luxembourg Wealth Study \citep{LWS:2020} are two examples of surveys released after performing MI, and used by sociologists publishing in AJS and ASR\@.

    \end{itemize}

    Despite its importance, congeniality should not be considered the sole guiding principle when defining imputation models.
    There are even cases where uncongeniality can improve on the efficiency of the standard complete data procedure, a phenomenon known as superefficiency (\citealp[pp. 544--546]{meng:1994}; \citealp[p. 481]{rubin:1996}; \citealp[pp. 217--218]{littleRubin:2002}).
    Furthermore, an imputation model that is congenial to a given analysis model may nevertheless fail to produce proper imputations.
    \citeauthor{rubin:1976} (\citeyear[pp. 584--585]{rubin:1976}) described the three conditions under which the distribution of the missingness is ignorable.
    The first of these conditions is that the missing data are missing at random (MAR), meaning that the probability of being missing is the same within groups defined by the observed data (i.e., conditioning on the observed data).
    When this condition is violated, standard MI can lead to biased parameter estimates, even if the analysis and imputation models are congenial.

    Meeting the MAR assumption requires specifying imputation models that include the variables that correlate with the missingness and the analysis model variables.
    Omitting such variables from the imputation model results in imputation under MNAR \citep[p. 339]{collinsEtAl:2001}.
    Applying standard MI under MNAR can lead to bias in the parameter estimates and can invalidate inferences involving the imputed variables \citep[pp. 341--343]{collinsEtAl:2001}.
    Therefore, including as many good predictors of the variables under imputation as possible in the imputation model is generally advisable.
    In this study, we focus on methods that assume MAR data.
    However, a considerable amount of research has been devoted to developing missing data treatments for MNAR data.
    We refer interested readers to \citet[pp. 287--328]{enders:2010}'s review of the two classes of MNAR models (i.e., selection models and pattern mixture models) and to \citet{littleZhang:2011}'s subsample ignorable multiple imputations: a method to obtain valid inferences with MI under MNAR under certain additional assumptions.

    We refer to all variables that are not targets of imputation, as potential auxiliary variables.
    This set of potential auxiliaries may include important predictors of missingness, variables that correlate with the imputation targets, and variables that are not useful for imputation.
    Discerning which of the potential auxiliary variables may be useful predictors in the imputation model can be a daunting task.
    Following an inclusive approach (i.e., including numerous auxiliary variables in the imputation model) reduces the chances of omitting important correlates of missingness, thereby making the MAR assumption more plausible (\citealp[pp. 826--827]{rubinEtAl:1995}; \citealp[p. 23]{schafer:1997}; \citealp{whiteEtAl:2011}; \citealp[p. 167]{vanBuuren:2018}).
    Furthermore, \cite{collinsEtAl:2001} showed that the inclusive strategy reduces estimation bias and increases efficiency.
    When designing broad and intermediate imputation models, the inclusive strategy can also grant congeniality with a wider range of analysis models.

    Although following the inclusive strategy may be beneficial for the imputation procedure, it is often infeasible to use all potential auxiliaries as predictors with standard imputation methods.
    Standard imputation methods, such as imputation under the normal linear model \citep[p. 68]{vanBuuren:2018}, face computational limitations in the presence of many predictors.
    For example, using traditional (unpenalized) regression models for the imputation model requires the number of predictors ($p$) in the imputation models to be smaller than the number of observed cases ($n$) to avoid mathematical singularity of the underlying system of equations \citep[p. 203]{jamesEtAl:2013}.
    As a result, imputers need to balance the benefits of the inclusive strategy with its computational limits.
    The large number of variables available in modern social scientific data sets makes the difficult step of deciding which predictors to include in the imputation models even more arduous.

    In addition to their size, other aspects of social surveys and other social scientific data can further complicate the task of specifying good imputation models.
    Sociologists and researchers working with large social surveys often want to estimate analysis models that use composite scores (i.e., aggregates of multi-item scales).
    When working with multi-item scales, the imputer needs to decide if variables should be imputed at the item level or at the scale level.
    When all a scale's items are usually missing or observed together, scale-level imputation can be effective \citep{mainzerEtAl:2021}.
    When item-level missing predominates, however, the literature generally suggests imputing multi-item scales at the item level \citep{vanBuuren:2010, gottschallEtAl:2012, eekhoutEtAl:2014}, but pursuing such a strategy can lead to increased dimensionality of the imputation models \citep{eekhoutEtAl:2018}.

    Furthermore, social surveys are often longitudinal, and it is usually most convenient to impute such a data structure in wide format \citep[p. 312]{vanBuuren:2018}.
    A wide data set has a single record for each unit, with observations made at subsequent time points coded as additional
    columns in the data set.
    As a result, long-running panel studies might easily induce large pools of potential auxiliary variables with which the imputer must contend.

\subsection{High-dimensional imputation}

    The factors discussed above---or combinations thereof---may result in \emph{high-dimensional} imputation problems wherein the pool of potential auxiliary variables is larger than the available sample size.
    Such high-dimensional problems preclude a straightforward application of MI and force researchers to choose which variables to include in the imputation model or otherwise regularize the imputation model.
    One possible solution to this problem is using high-dimensional prediction models as the imputation model.
    When we say ``high-dimensional prediction'', we are referring to the branch of statistical prediction concerned with improving prediction in situations where the number of predictors is larger than the number of observed cases (the so-called $p > n$ problem).
    Recent developments in high-dimensional imputation techniques leverage high-dimensional prediction methodology to offer opportunities for embracing an inclusive strategy while substantially diminishing its downsides.

    MI has been combined with high-dimensional prediction models in algorithms that use shrinkage methods \citep{zhaoLong:2016, dengEtAl:2016} and dimensionality reduction \citep{songBelin:2004, howardEtAl:2015} to avoid the obstacles of an inclusive strategy.
    Tree-based imputation strategies \citep{burgetteReiter:2010, dooveEtAl:2014} also have the potential to overcome the computational limitations of the inclusive strategy.
    The nonparametric nature of decision trees bypasses the identification issues most parametric methods face in high-dimensional contexts.
    To the best of our knowledge, no study to date has directly compared the performance of the various high-dimensional MI methods recommended in the literature.

\subsection{Scope of the current project}

    The goal of this project was to compare how different high-dimensional MI (HD-MI) methods fare when imputing data sets with many variables.
    In particular, we were interested in the types of imputation problems that may arise in large social scientific data sets.
    Such data sets do not need to be strictly high-dimensional to be \textit{too large} for standard MI routines.
    Even in low-dimensional settings (i.e., $n > p$), including too many auxiliary variables in the imputation model can bias analysis model estimates and lead to convergence problems and other computational issues \citep{hardtRainer:2012}.
    The high-dimensional imputation approaches we compared in this project can be used to simplify the process of specifying a good imputation model in both high- and low-dimensional problems.

    We compared seven state-of-the-art HD-MI algorithms in terms of their ability to support statistically valid analyses.
    We chose these techniques because they stood out as the most promising candidates in our review of the HD-MI literature.
    The comparison was based on two numerical experiments: a Monte Carlo simulation study and a resampling study using Wave 5 of EVS\@.
    The simulation study allows us to compare the imputation methods in an artificial scenario with maximum experimental control.
    In a simulation study, we are able to precisely manipulate data features to match our experimental goals because we define the population model.
    However, the variables in a simulation study are usually sampled from simple multivariate distributions with regular, unrealistic mean and covariance structures.
    The resampling study allows us to shed the artifice of the simulation study and compare the methods using real social scientific data.
    EVS is a large-scale, cross-national survey on human values administered in almost 50 countries across Europe.
    The EVS data contain both numerical and categorical variables associated via a complicated, heterogeneous covariance structure.
    Performing a resampling study on this data set allows us to estimate bias and coverage in a more ecologically valid---albeit still somewhat artificial---scenario than is possible with a Monte Carlo simulation study.\footnote{We chose to use the EVS data for a resampling study rather than an applied example because there is no ground truth in an applied example. The resampling study offers much of the ecological validity of an applied example with the added benefit of supporting the same types of generalizable conclusions provided by a simulation study.}

    The imputation techniques we compared are best suited to data-driven imputation with an intermediate or broad scope.
    The potential benefits of HD-MI methods lie in the automatic imputation model specification that these techniques offer.
    Therefore, we focused on data-driven imputation tasks where the objective is accommodating a wide range of analysis models.
    However, the techniques we compared do not exclude the possibility of specifying more narrowly scoped imputation models.
    With little tweaking, one can always force specific variables into the imputation model.

    In what follows, we first introduce the missing data treatments that we compared in our study.
    Then, we present the methodology and results of the two numerical experiments, we discuss the implications of the results for applied researchers, and we provide recommendations.
    We conclude by discussing the limitations of the study and suggesting future research directions.
    \section{Imputation methods and algorithms}\label{sec:methods}

	We use the following notation: scalars, vectors, and matrices are denoted by italic lowercase, bold lowercase, and bold uppercase letters, respectively.
	A scalar belonging to an interval is indicated by $s_1 \in [s_2, s_3]$, while a scalar taking the values in a set is represented as $s_1 \in \{s_2, s_3\}$.
	We use the scope resolution operator, ::, to designate a function provided by a specific software package.
	So, for example, \emph{mice::quickpred()} represents the \emph{quickpred()} function provided by the \emph{mice} package.

	Consider an $n \times p$ data set, $\bm{Z}$, comprising variables $\bm{z}_1$, $\bm{z}_2$, \dots, $\bm{z}_p$.
	Assume that the first $t$ variables of $\bm{Z}$ have missing values and that these $t$ variables are the targets of imputation.
	Denote the columns of $\bm{Z}$ containing $\bm{z}_1$ to $\bm{z}_t$ as the $n \times t$ matrix, $\bm{T}$.
	The remaining $(p-t)$ columns of $\bm{Z}$ contain variables that are not targets of imputation.
	These variables constitute a pool of \emph{potential} auxiliary variables that could be used to improve the imputation procedure.
	Let $\bm{A}$ be a $n \times (p-t)$ matrix denoting this set of potential auxiliary variables and write $\bm{Z}$ as $\bm{Z} = (\bm{T}, \bm{A})$.
	For a given $\bm{z}_j$, with $j = (1, \dots, p)$, denote its observed and missing components as $\bm{z}_{j, obs}$ and $\bm{z}_{j, mis}$, respectively.
	Let $\bm{Z}_{-j} = (\bm{z}_1, \dots, \bm{z}_{j-1}, \bm{z}_{j+1}, \dots, \bm{z}_{p})$ be the collection of $p-1$ variables in
	$\bm{Z}$ excluding $\bm{z}_j$.
	Denote by $\bm{Z}_{-j, obs}$ and $\bm{Z}_{-j, mis}$ the components of $\bm{Z}_{-j}$ corresponding to the data units in $\bm{z}_{j, obs}$ and $\bm{z}_{j, mis}$, respectively.

\subsection{Multiple imputation by chained equations}

	Assume that $\bm{Z}$ is the result of $n$ random samples from a multivariate distribution defined by an unknown set of parameters $\bm{\theta}$.
	The multiple imputation by chained equations (MICE) approach obtains the posterior distribution of $\bm{\theta}$ by sampling iteratively from conditional distributions of the form $P(\bm{z}_{1}|\bm{Z}_{-1}, \bm{\theta}_{1}), \dots, P(\bm{z}_{t}|\bm{Z}_{-t}, \bm{\theta}_{t})$, where $\bm{\theta}_{1}, \dots, \bm{\theta}_{t}$ are imputation model parameters specific to the conditional distributions of each variable with missing values.

	More precisely, the MICE algorithm takes the form of a Gibbs sampler\footnote{Technically, the MICE algorithm is only a true Gibbs sampler when there exists a valid joint distribution for all targets of imputation.
	When such a joint distribution does not exist, the MICE algorithm is still valid, but it is not a true Gibbs sampler.}.
	At the $m$th iteration $(m = 1, \dots, M)$, samples are drawn for the $j$th target variable ($j = 1, \dots, t$) from the following distributions:

	\begin{align}
		\hat{\bm{\theta}}_{j}^{(m)} &\sim
		p(\bm{\theta}_j | \bm{z}_{j, obs}, \dot{\bm{Z}}_{-j, obs}^{(m)})
		\label{eq:pd}\\
		\bm{z}_{j, mis}^{(m)} &\sim
		p(\bm{z}_{j, mis} | \dot{\bm{Z}}_{-j, mis}^{(m)}, \hat{\bm{\theta}}_{j}^{(m)})
		\label{eq_ppd},
	\end{align}
	where $\hat{\bm{\theta}}_{j}^{(m)}$ and $\bm{z}_{j, mis}^{(m)}$ are draws from the parameter's full conditional posterior distribution \eqref{eq:pd} and the missing data posterior predictive distribution \eqref{eq_ppd}, respectively.
	$\dot{\bm{Z}}_{-j, obs}^{(m)}$ and $\dot{\bm{Z}}_{-j, mis}^{(m)}$ are subsets of the variables in $\bm{Z}_{-j}^{(m)}$ (potentially every variable in $\bm{Z}_{-j}^{(m)}$).
	These subsets are chosen by the imputer to act as predictors in the elementary imputation model for $\bm{z}_j$.
	After convergence, $d$ sets of values are sampled from \eqref{eq_ppd} and used as imputations.
	Any analysis model can then be estimated on each of the $d$ completed data sets, and the parameter estimates can be pooled using Rubin's rules \citep{rubin:1987}.

	In the following, we describe all the missing data treatments we compared in this study.
	First, we describe the seven \emph{high-dimensional MICE strategies} we compared in this study.
	They follow the general MICE framework, but they differ in which elementary imputation methods they use to define equations \eqref{eq:pd} and \eqref{eq_ppd}.
	Second, we describe three \emph{benchmark mice strategies}, which are well-established approaches in the field of sociology and the missing data treatment literature.
	Finally, we describe two \emph{benchmark non-MI strategies}, which are important baselines of comparisons that do not rely on imputation.

\subsection{High-dimensional MICE strategies}

\subsubsection{MICE with step-forward selection}

	A linear regression model is the standard univariate imputation model for MICE\@.
	However, ordinary linear regression (OLS) faces computational limitations when applied to data sets with many predictors.
	If $n$ is not much larger than $p$, the regression estimates will have large variances, and, if $p>n$, there is no unique solution for the regression coefficients.
	Researchers have been studying model-building strategies to overcome these limitations for decades (e.g., \citealp{dempsterEtAl:1977b}).
	One of these strategies, known as forward stepwise subset selection \citep{efroymson:1966}, has been implemented in the popular imputation software IVEware \citep{raghunathanEtAl:2002}.
	We refer to this method as MI Step-Forward (MI-SF).

	Forward selection identifies the subset of the predictors that are most related to the dependent variables by iteratively evaluating the improvement in fit contributed by including each additional predictor.
	Starting with an empty imputation model, MI-SF iteratively adds the variable that most increases the model-explained variance.
	New predictors are added as long as the additional proportion of variance they explain exceeds a specified threshold value $R^2_{min}$.
	As a result, MI-SF ensures that the predictors included in equation \eqref{eq:pd} must explain some non-trivial proportion variability in the variable under imputation.
	The value of $R^2_{min}$ used in the MI-SF algorithm is fixed across iterations, but the imputation model for every variable might change between iterations.

\subsubsection{MICE with a fixed ridge penalty}

  	So-called \emph{shrinkage methods} represent an alternative to subset selection (\citealp[see][pp. 62--79]{hastieEtAl:2009} for a review.)
	These methods address the computational problems caused by large numbers of predictors by shrinking the estimated coefficients toward zero.
	Ridge regression \citep{hoerlKennard:1970} is a common shrinkage method that imposes a penalty during model estimation to shrink the regression slopes toward zero and allow a large number of predictors to be included in the model, while still controlling the variance of the estimates.
	When applied to the imputation model in MICE, a ridge penalty allows a more inclusive auxiliary variable strategy.

	MICE with a fixed ridge penalty uses the Bayesian normal linear model described by \citet[][p. 68, algorithm 3.1]{vanBuuren:2018} as the univariate imputation method.
	We refer to this method as Bayesian Ridge (BRidge).
	In this approach, the sampling of each $\hat{\bm{\theta}}_{j}^{(m)}$ in equation \eqref{eq:pd} relies on inverting the cross-products matrix of $\dot{\bm{Z}}_{-j, obs}^{(m)}$\footnote{When estimating ridge regression coefficients, predictors are centered and scaled to have unit variance}.
	Adding a positive constant (the ridge penalty, $\kappa$) to the diagonal of the cross-product matrix stabilizes this inversion.
	Indeed, if $p > n$, sufficiently large values of $\kappa$ will facilitate inversion of the cross-products matrix and induce a unique (albeit biased) solution for the regression coefficients.

	In BRidge, every variable in $\bm{Z}_{-j}$ is used as a predictor in the imputation model, and the ridge penalty is the only precaution taken to address a large number of predictors.
	The value of $\kappa$ is usually chosen to be close to zero (e.g. $\kappa = 0.0001$), because values larger than $0.1$ may introduce excessive systematic bias \citep[p. 68]{vanBuuren:2018}.
	However, larger values of $\kappa$ may be necessary to adequately stabilize the estimation in certain scenarios.
	In the present work, we chose the value of $\kappa$ by means of cross-validation.

\subsubsection{Direct use of regularized regression}\footnote{The DURR approach is now implemented in the R package \emph{mice} via the \textit{mice::mice.impute.lasso.norm()} and \textit{mice::mice.impute.lasso.logreg()} elementary imputation methods.}

	Lasso regression \citep[\emph{least absolute shrinkage and selection operator}; ][]{tibshirani:1996} is another popular shrinkage method.
	Unlike ridge regression, the lasso penalty achieves both shrinkage and automatic variable selection (whereas ridge does not exclude any variables).
	The extent of the lasso penalization depends on a tuning parameter, $\lambda$, which is selected from a set of possible values by means of cross-validation.
	For sufficiently large values of $\lambda$, lasso will force some coefficient estimates to be exactly zero thereby excluding the associated predictors from the fitted model.
	When applied to an imputation model, lasso will automatically select which predictors enter the imputation model.
	\cite{zhaoLong:2016} and \cite{dengEtAl:2016} used lasso regression as the univariate imputation model in a MICE algorithm to impute high-dimensional data and referred to this approach as \emph{direct use of regularized regression} (DURR).

	At iteration $m$, for a target variable $\bm{z}_j$, DURR replaces equations \eqref{eq:pd} and \eqref{eq_ppd} with the following two steps:
	\begin{enumerate}

		\item Generate a bootstrap sample $\bm{Z}^{*(m)}$ by sampling with replacement from $\bm{Z}^{(m)}$, and train a regularized linear regression model (such as lasso regression) with $\bm{z}_{j,obs}^{*(m)}$ as outcome and $\bm{Z}_{-j,obs}^{*(m)}$ as predictors\footnote{As with ridge regression, predictors are centered and scaled to have unit variance.}.
		This produces a set of parameter estimates (regression coefficients and error variance), $\hat{\bm{\theta}}_{j}^{(m)}$, that can be viewed as a sample from equation \eqref{eq:pd}.

		\item Use $\bm{Z}^{(m)}_{-j, mis}$ and $\hat{\bm{\theta}}_{j}^{(m)}$ to predict $z_{j,mis}$, and obtain draws from the posterior predictive distribution of the missing data as in equation
		\eqref{eq_ppd}.

	\end{enumerate}
	Hence, at every iteration, each elementary imputation model is estimated as a lasso regression, and uncertainty regarding the parameter values is included by bootstrapping.

	In high-dimensional cases, lasso selects at most $n$ predictors \citep{zouHastie:2005}.
	So, when using lasso for imputation, no elementary imputation model will contain more predictors than the number of observed cases on the corresponding outcome.
	\cite{dengEtAl:2016} compared lasso with the elastic net---which does not have this restriction---for high-dimensional MI, but they did not find evidence to favor the elastic net over lasso.
	Lasso is also computationally simpler than the elastic net because lasso only has one tuning parameter to estimate whereas the elastic net has two.
	Therefore, we chose to implement DURR with lasso as the regularization method.

\subsubsection{Indirect use of regularized regression}\footnote{The IURR approach is now implemented in the R package \emph{mice} via the \textit{mice::mice.impute.lasso.select.norm()} and \textit{mice::mice.impute.lasso.select.logreg()} elementary imputation methods.}

	While DURR simultaneously performs model regularization and parameter estimation in equation \eqref{eq:pd},
	the \emph{indirect use of regularized regression} \citep[IURR; ][]{zhaoLong:2016, dengEtAl:2016} algorithm uses regularized regression exclusively for variable selection.
	The selected variables are then used as predictors in the imputation models of a standard MI procedure.

	At iteration $m$, the IURR algorithm performs the following steps for each target variable, $\bm{z}_j$:
	\begin{enumerate}

		\item Fit a linear regression model using a regularized method that does variable selection (e.g., lasso).
		Take $\bm{z}_{j,obs}$ as the dependent variable and $\bm{Z}_{-j,obs}^{(m)}$ as the predictors (unlike DURR, IURR uses the original data, not a bootstrap sample).
		The regression coefficients that are \emph{not} shrunk to 0 define the active
		set of variables that will be used as predictors in the actual imputation model (i.e., the variables in $\dot{\bm{Z}}_{-j}^{(m)}$). \label{varSelectStep}
		\item Obtain the maximum likelihood estimates of the regression coefficients and the error variance from the linear
		regression of $\bm{z}_{j,obs}$ onto the active set of predictors defined in step \ref{varSelectStep}.
		Then, sample new values of these parameters from a multivariate normal distribution
		parameterized by the MLEs\footnote{The sampling notation is the same used by \cite{dengEtAl:2016}.}:
		\begin{equation}\label{eq:mlepd}
		(\hat{\bm{\theta}}_{j}^{(m)}, \hat{\sigma}_{j}^{(m)}) \sim N(\hat{\bm{\theta}}_{MLE}^{(m)},
		\hat{\bm{\Sigma}}_{MLE}^{(m)})
		\end{equation}
		so that equation \eqref{eq:mlepd} corresponds to equation \eqref{eq:pd} in the general MICE framework.
		\item Impute $\bm{z}_{j,mis}$ by sampling from the posterior predictive distribution based
		on $\dot{\bm{Z}}_{-j,mis}^{(m)}$ and the parameters' posterior draws, $(\hat{\bm{\theta}}_{j}^{(m)},
		\hat{\sigma}_{j}^{(m)})$.
	\end{enumerate}

	DURR uses regularized regression to directly obtain $\hat{\bm{\theta}}_{j}^{(m)}$, a procedure that inherently induces estimation bias.
	Compared to DURR, IURR separates the variable selection step, which involves using the biasing penalty term, from the sampling of the imputation model parameters.
	Assuming the variable selection step does not exclude any important
        predictors, the two-step approach of IURR could outperform DURR by using unbiased estimates of $\hat{\bm{\theta}}_{MLE}^{(m)}$ and $\hat{\bm{\Sigma}}_{MLE}^{(m)}$ to
        define the posterior distributions of the imputation model parameters.
	IURR effectively establishes a data-driven decision rule to select imputation model predictors while avoiding the direct involvement of the biasing penalty in the simulation of a random draw from Equation \eqref{eq:pd}.

\subsubsection{MICE with Bayesian lasso}

	\cite{zhaoLong:2016} proposed the MICE with Bayesian Lasso imputation algorithm (BLasso), an MI procedure that uses the Bayesian lasso as its elementary imputation method: MICE with Bayesian lasso (BLasso).
	A Bayesian lasso model is a regular Bayesian multiple regression model with informative priors on the slope coefficients that allow interpreting the mode of the slopes' posterior distribution as lasso estimates
	\citep{parkCasella:2008, hans:2009}.
	Following \cite{zhaoLong:2016}, we used the Bayesian lasso specification given by \cite{hans:2010}.
	Given data with a sample size, $n$, a dependent variable, $\bm{y}$, and a set of predictors, $\bm{X}$, the Bayesian lasso model has the following form.

	\begin{align}
		p(\bm{y}|\bm{\beta}, \sigma^2, \tau) &= \textrm{N}(\bm{y}|\bm{X\beta}, \sigma^2\bm{I}_n) \label{eqn:dens} \\
		p(\beta_j|\tau, \sigma^2, \rho) &=
		(1 - \rho) \delta_0 (\beta_j) +
		\rho \left( \frac{\tau}{2\sigma} \right) \times
		\textrm{exp} \left( \frac{-\tau \norm{\beta_j}_1}{\sigma} \right) \label{eqn:bprior} \\
		\sigma^2 &\sim \textrm{Inverse-Gamma}(a, b) \label{eqn:sigprior} \\
		\tau &\sim \textrm{Gamma}(r, s) \label{eqn:tauprior} \\
		\rho &\sim \textrm{Beta}(g, h) \label{eqn:rhoprior}
	\end{align}
	Equation \eqref{eqn:dens} represents the density function of a multivariate normal random variable
	with mean $\bm{X\beta}$ and covariance matrix $\sigma^2\bm{I}_n$ evaluated at $\bm{y}$.
	Equation \eqref{eqn:bprior} is the mixture prior distribution for the regression coefficients $\beta_j$ proposed by \cite{hans:2010}.
	This formulation differs from the classical Bayesian lasso prior proposed by \cite{parkCasella:2008} because of the presence of the sparsity parameter, $\rho$ (\citealp[pp. 655--656]{leySteel:2009};~\citealp[pp. 2592]{scottBerger:2010}), and the point mass at zero, $\delta_0 (\beta_j)$.
	Finally, equations \eqref{eqn:sigprior} to \eqref{eqn:rhoprior} represent hyper priors for the residual variance,
	$\sigma^2$, the penalty parameter, $\tau$, and the sparsity parameter, $\rho$.
	Our implementation of BLasso imputation replaced equation \eqref{eq:pd} with the BLasso model defined by equations \eqref{eqn:dens} to \eqref{eqn:rhoprior} with $\bm{y} = \bm{z}_{j,obs}$ and $\bm{X} = \bm{Z}_{-j,obs}$.

	The R code used to perform the BLasso imputation was based on the R Package \emph{blasso} \citep{blasso} and can be found in the code repository for this article \citep{costantini:2023d}.
	For a detailed description of the Bayesian lasso MI algorithm in a univariate missing data context see \cite{zhaoLong:2016}.

\subsubsection{MICE with PCA}

	By extracting principal components (PCs) from the set of potential auxiliary variables, $\bm{A}$, the \emph{MICE with PCA} (MI-PCA) method summarizes the information contained in $\bm{A}$ with just a few components.
	These PCs can then be used as predictors in a standard, low-dimensional application of MICE\@.
	The MI-PCA procedure can be summarized as follows:
	\begin{enumerate}

		\item Extract the first PCs that cumulatively explain the desired proportion of the variance in the set of potential auxiliary variables, $\bm{A}$\footnote{The columns of $\bm{A}$ are centered and scaled to have unit variance.}, and collect these components in a new matrix, $\bm{A}'$;

		\item Replace $\bm{A}$ in $\bm{Z}$ with $\bm{A}'$ to obtain $\bm{Z}' = (\bm{T}, \bm{A}')$;

		\item Use the standard MICE algorithm with a Bayesian normal linear model and no ridge penalty to obtain multiply imputed data sets from $\bm{Z}'$.

	\end{enumerate}

	The MI-PCA method was inspired by \cite{howardEtAl:2015} and the \emph{PcAux} R package \citep{PcAux}.
	For this study, we used the R function \emph{stats::prcomp()} to perform the PCA estimation via truncated singular value decomposition.
	Hence, $p > n$ data are not a problem.
	When $\bm{A}$ has more columns than rows, \emph{prcomp()} will simply extract a maximum of $n$ components.
	
\subsubsection{MICE with classification and regression trees}

	MICE with classification and regression trees \citep[MI-CART;][]{burgetteReiter:2010} is a MICE algorithm that uses classification and regression trees (CART) as the elementary imputation method.
	Given an outcome variable $\bm{y}$ and a set of predictors $\bm{X}$, CART is a nonparametric recursive partitioning technique that models the relationship between $\bm{y}$ and $\bm{X}$ by sequentially splitting observations into subsets of units with relatively more homogeneous $\bm{y}$ values.
	At every splitting stage, the CART algorithm searches through all variables in $\bm{X}$ to find the best binary partitioning rule to predict $\bm{y}$.
	The resulting collection of binary splits can be visually represented by a decision tree structure where each terminal node (or \emph{leaf}) represents the conditional distribution of $\bm{y}$ for units that satisfy the splitting rules.

	For each $\bm{z}_j$, the $m$th iteration of MI-CART proceeds as follows:

	\begin{enumerate}

		\item Train a CART model to predict $\bm{z}_{j, obs}$ from the corresponding $\bm{Z}_{-j, obs}^{(m)}$.

		\item Assign each element of $\bm{z}_{j,mis}$ to a terminal node by applying the splitting rules from the fitted CART model to $\bm{Z}^{(m)}_{-j,mis}$.

		\item Create imputations for each element of $\bm{z}_{j,mis}$ by sampling from the pool of $\bm{z}_{j, obs}$ in the terminal node containing $\bm{z}_{j,mis}$.
		This procedure corresponds to sampling from the missing data posterior predictive distribution in Equation \eqref{eq_ppd}.

	\end{enumerate}

	This approach does not consider uncertainty in the imputation model parameters since the tree structure is not perturbed between iterations.
	Therefore, MI-CART cannot produce proper imputations in the sense of \citet{rubin:1986}.
	The implementation of MI-CART used in this paper corresponds to the one presented by \citet[][p. 95, algorithm 1]{dooveEtAl:2014} and the \emph{impute.mice.cart()} function from the \emph{mice} package.

	CART searches for the best splitting criterion one variable at a time.
	As a result, $p > n$ does not pose the same computational limitations that plague methods based on linear regression.
	More variables can increase estimation times but will not result in computational obstructions.

\subsubsection{MICE with random forests}

	MICE with random forests (MI-RF) is a MICE algorithm that uses random forests as the elementary imputation method.
	The random forest algorithm \citep[e.g., ][p. 588]{hastieEtAl:2009} entails fitting many decision trees (e.g., CART models) to subsamples of the original data.
	These subsamples are derived by resampling rows with replacement and sampling subsets of columns without replacement.
	The random forest algorithm results in an ensemble of fitted decision trees that generate a sample of predictions for each outcome value.
	Consequently, random forests often demonstrate better prediction performance than individual trees by reducing the variance of the estimated prediction function.

	For each $\bm{z}_j$, the $m$th iteration of MI-RF proceeds as follows:
	\begin{enumerate}
		\item Generate $k$ bootstrap samples from $\bm{Z}_{-j,obs}$.
		\item Use these bootstrap samples to fit $k$ single trees predicting $\bm{z}_{j,obs}$ from a random subset of the variables in $\bm{Z}_{-j,obs}$.
		\item Generate a pool of $k$ terminal nodes for each element of $\bm{z}_{j,mis}$ by applying the splitting rules from each of the $k$ fitted trees to the appropriate columns of $\bm{Z}_{-j,mis}$.
		\item Create imputations for each element of $\bm{z}_{j,mis}$ by sampling from the $\bm{z}_{j,obs}$ contained in the pool of terminal nodes defined above.
	\end{enumerate}

	Bootstrapping and random input selection introduce uncertainty regarding the imputation model parameters (i.e., the tree structure), as required by a proper MI procedure.
	For more details on the MI-RF algorithm, see \citet[p. 103]{dooveEtAl:2014}.
	To perform MICE with random forests we used the R function \emph{mice::impute.mice.rf()}.
	As with CART, the random forests algorithm is not subject to computational limitations in high-dimensional problems because random forests simply aggregate a collection of univariate decision trees.

\subsection{Benchmark MICE strategies}

\subsubsection{MICE with quickpred}

	A simple way to select predictors for an imputation model is to include variables that relate to the nonresponse or explain a considerable amount of variance in the targets of imputation.
	One popular implementation of this idea is to select as predictors those variables whose association with the variables under imputation, or their response indicators, exceeds some threshold.
	This selection strategy was proposed by \citet{vanBuurenEtAl:1999} and has been implemented in the \emph{quickpred} function provided by the popular R package \emph{mice} \citep[p. 267]{vanBuuren:2018}.
	We refer to this approach as MI-QP\@.
	As both an intuitive, pragmatic option and the default method of selecting predictors in one of the most popular MI software packages, MI-QP represents an important benchmark against which to compare the performance of the more theoretically sound approaches described above.

	The MI-QP approach has two main drawbacks. First, selecting predictors based on their correlations with the targets of imputation and the associated response indicators can still select collinear, redundant predictors.
	If one predictor is highly correlated with another and with a variable under imputation, both will be selected.
	Second, when applied to $p > n$ scenarios, MI-QP is not guaranteed to select fewer predictors than observations available for a given imputation model.
	As a result, MI-QP often needs to be augmented by other techniques to address collinearity and linear dependencies in the data.

\subsubsection{MICE with analysis model variables as predictors}

	According to our review of the articles published in AJS and ASR, a common approach to address the large number of possible predictors is to use only the analysis model variables in the imputation model.
	We refer to this approach as MI-AM\@.
	Consider a researcher working with EVS data who wants to estimate a linear model by regressing one item on 10 others afflicted by non-response. 
	The MI-AM imputation strategy would imply using only these 11 variables in the imputation models, instead of manually searching all of the 250 variables contained in the survey for meaningful imputation predictors.
	
	The MI-AM strategy ensures the congeniality of the analysis and imputation models.
	Furthermore, as long as the analysis model does not include more variables than the number of observed cases, MI-AM is not affected by the dimensionality of the data.
	However, by following this strategy, any MAR predictors that are not part of the analysis model will be excluded from the imputation. 
	In such cases, the MAR assumption is violated, and the missingness is MNAR.

\subsubsection{Oracle MICE}

	As hinted by the previous two approaches, the MI literature recommends following three principles to decide which predictors to include in the imputation models \citep[p. 168]{vanBuuren:2018}:
	\begin{enumerate}

		\item Include all variables that are part of the analysis model(s).

		\item Include all variables that are related to the nonresponse.

		\item Include all variables that are correlated with the targets of imputation.

	\end{enumerate}

	In practice, the first criterion can be met only if the analysis model is known before imputation, which is not always true.
	Furthermore, researchers can never be sure that the second criterion is entirely met, as there is no way to know exactly which variables are responsible for missingness.
	However, with simulated data, we know which variables define the response model.
	The oracle MICE approach (MI-OR) is an ideal specification of the MICE algorithm that uses this knowledge to include only the relevant predictors in the imputation models.
	As such, this method cannot be used in practice, but it provides a useful reference point for the desirable performance of an MI procedure.
	The MI-OR imputations were generated using the Bayesian normal linear model as the univariate imputation method.
	
\subsection{Non-MI strategies}

\subsubsection{Complete case analysis}

	By default, most data analysis software either fails in the presence of missing values or defaults to analyzing only the complete cases \citep{R:2020, pandas:2020}.
  	As the default behavior of most statistical software, complete cases analysis (CC) remains a popular missing data treatment in the social sciences \citep{peughEnders:2004, littleEtAl:2013}.
	CC can also be a useful approach in certain scenarios \citep{whiteCarlin:2010}.
	For example, when the analysis model is a linear regression of $y$ onto a set of predictors, $\bm{X}$, CC yields valid inferences if the missingness depends only on $\bm{X}$ and not on $y$ (\citealp[p. 43]{littleRubin:2002}; \citealp{littleZhang:2011}).
	However, even in this case, CC can be inefficient as it uses a reduced sample size compared to what could be used through proper imputation (\citealp[p. 42]{littleRubin:2002}; \citealp{schaferGraham:2002}).
	Furthermore, unless the data are missing completely at random (MCAR), CC can bias parameter estimates (\citealp[p. 8]{rubin:1987}, \citealp{schaferGraham:2002}).
  	Nevertheless, the continued popularity of CC makes it an important benchmark method.

\subsubsection{Gold standard}

	We also estimated the analysis models directly on the fully observed data before imposing any missing values.
	In the following, we refer to the results obtained in this fashion as the gold standard (GS).
	These results represent the counterfactual analysis that would have been performed if there had been no missing data.
    \section{Simulation study}

We investigated the performance of the methods described above with a Monte Carlo simulation study.
Following a similar procedure to that employed by
\cite{collinsEtAl:2001}, we generated $S = 1000$ samples of $n = 200$
units while varying two experimental factors: the number of variables in
the data set, $p \in \{50, 500\}$, and the proportion of missing cases on
each of the incomplete variables, $pm \in \{0.1, 0.3\}$.
Table \ref{tab:condExp1} summarizes the four resulting crossed conditions.

We chose the values of $n$ and $p$ to reflect extreme dimensionality situations that would tease apart the relative strengths and weaknesses of the imputation methods considered here.
Nonetheless, we selected these values to be somewhat plausible for real-world social scientific studies.
Consider, for example, that a typical EVS wave has around 55,000 observations and 250 items in its questionnaire.
Therefore, data structures similar to those in both our low- and high-dimensional conditions could arise by taking reasonable subsets of EVS data (potentially over several waves).
As for the levels of $pm$, we chose the lower level to match the 10\% of missing cases that is typical of variables in EVS data.
We also included a more extreme level to create more challenging---but still realistic---conditions for the imputation methods.
For every iteration, we imposed missing values on six target items, and then we used all missing data treatment methods described above to obtain estimates of the means, variances, and covariances of these incomplete variables.

\begin{table}
	\centering
	\begin{tabular}{l c c c c }
		\toprule
		condition & label            & n   & p   & pm  \\
		\midrule
		1         & low-dim-low-pm   & 200 & 50  & 0.1 \\
		2         & high-dim-low-pm  & 200 & 500 & 0.1 \\
		3         & low-dim-high-pm  & 200 & 50  & 0.3 \\
		4         & high-dim-high-pm & 200 & 500 & 0.3 \\
		\bottomrule
	\end{tabular}
	\caption{Summary of conditions for Experiment 1. Low-dim (high-dim) represent conditions where the number of predictors
		is smaller (larger) than the number of observations available. Low-pm (high-pm) represent conditions where the
		proportion of missing values is low (high)}
	\label{tab:condExp1}
\end{table}

\subsection{Simulation study procedure}\label{sec:simstudyproc}

\subsubsection{Data generation}

At every replication, a data matrix $\bm{Z}_{n \times p}$ was generated according to a multivariate normal model with means equal to 5 and unit variances.
The distribution was centered around 5 as typical 10-point numerical items in the EVS data set have means around 5.
After sampling the data, all variables were rescaled to have a variance of approximately 5, which reflects the typical size of the variance of 10-point items in the EVS data.
For the correlation structure, we defined three blocks of variables based on three strengths of association: strong, weak, and none.
The first five variables were strongly correlated ($\rho = 0.6$) among themselves; variables 6 to 10 were weakly correlated ($\rho = 0.3$) with the first 5 variables and among themselves; the remaining $p-10$ variables were uncorrelated with any other variable in the data set.
Of course, real survey data have more complex correlation structures than what we defined for this study.
However, when specifying imputation models for survey data, the main challenge is often finding a few important auxiliary variables in a large collection of possible predictors.
We defined the population correlation matrix with the three-block structure described above to replicate this type of situation in an experimentally unequivocal way.\footnote{We used fixed correlation levels instead of varying the correlation values (e.g., to manipulate collinearity) for the same reason.
	Varying the strengths of association within blocks would have diluted this key feature of our population model.}.

\subsubsection{Missing data imposition}\label{sub_missing}

Missing values were imposed on six of the items in $\bm{Z}$: three variables in the block of highly correlated variables $\{\bm{z}_1, \bm{z}_2, \bm{z}_3\}$ and three in the block of lowly correlated variables $\{\bm{z}_6, \bm{z}_7, \bm{z}_8\}$.
Item nonresponse was imposed by sampling from a Bernoulli distribution with individual probabilities of nonresponse defined by
\begin{equation} \label{eq:rm}
	p_{miss} = p(z_{i,j} = miss | \tilde{\bm{Z}}) = \frac{ exp(\gamma_0 + \tilde{\bm{z}}_{i}\bm{\gamma}) }
	{ 1 + exp(\gamma_0 + \tilde{\bm{z}}_{i}\bm{\gamma}) },
\end{equation}
where $z_{i,j}$ is the $i$th subject's response on $\bm{z}_j$, $\tilde{\bm{z}}_{i}$ is a vector of responses to the set of missing data predictors for the $i$th individual, $\gamma_0$ is an intercept parameter, and $\bm{\gamma}$ is a vector of slope parameters.
$\tilde{\bm{Z}}$ was specified to include two fully observed variables from the strongly correlated set and two from the weakly correlated set $\{\bm{z}_4, \bm{z}_5, \bm{z}_9, \bm{z}_{10}\}$.
Therefore, the probability of nonresponse for a variable depended on variables present in the data, but never on the variable itself.
As a result, when the elements of $\tilde{\bm{Z}}$ are included as predictors in the MI procedures, the MAR assumption is satisfied.
All slopes in $\bm{\gamma}$ were fixed to 1, while the value of
$\gamma_0$ was chosen through numerical optimization to produce the desired proportion of missing values\footnote{The pseudo R-squared for the logistic regression of the missing value indicator on the predictors of missingness was around 15\%.
	The AUC for the logistic regression was around 0.72.
}.

\subsubsection{Imputation}

We generated ten imputed data sets by imputing the missing values with all methods described in the preceding section.
To evaluate the convergence of the imputation models, we ran ten replications of the \emph{high-dim-high-pm} condition and generated trace plots of the imputed values' means.
The implementation of MI-SF in IVEware does not provide trace plots. 
Therefore, we plotted the distributions of the imputed values across 30 imputation chains against the observed data at iterations 1, 5, 10, 20, 40, 80, 160, 240, and 320.
Based on the information provided by density and trace plots, we considered all of the imputation algorithms to have converged after 50 iterations.

IVEware does not offer any data-driven procedure for selecting $R^2_{min}$; and the IVEware authors recommend comparing results obtained with different $R^2_{min}$ values.
To optimize the performance of MI-SF, we tuned this parameter with a cross-validation procedure. 
We applied MI-SF with different $R^2_{min}$ values (i.e., $10^{-1}, 10^{-2}, \dots, 10^{-7}$), and we selected the value that resulted in the smallest average fraction of missing information \citep[FMI;][ eq. 3.1.10]{rubin:1987} across the analysis model parameters.
The same cross-validation strategy was used to choose the value of the ridge penalty in the BRidge algorithm.
We considered the values $10^{-1}, 10^{-2}, \dots, 10^{-8}$ as candidates for the BRidge penalty parameter.

Both IURR and DURR could have been implemented with a variety of penalties (e.g., lasso, \citealp{tibshirani:1996}; elastic net, \citealp{zouHastie:2005}; adaptive lasso, \citealp{zou:2006}).
In this study, we used lasso as it is computationally efficient, and it performed well for imputation in \cite{zhaoLong:2016} and \cite{dengEtAl:2016}.
A 10-fold cross-validation procedure was used at every iteration of DURR and IURR to choose the penalty parameter.
To maintain consistency with previous research, we specified the BLasso hyper-parameters in equations \eqref{eqn:sigprior}, \eqref{eqn:tauprior}, and \eqref{eqn:rhoprior} as in \cite{zhaoLong:2016}: $(a,b)=(0.1, 0.1)$, $(r,s)=(0.01, 0.01)$, and $(g,h)=(1,1)$.
For the MI-PCA algorithm, the set of possible auxiliary variables in $\bm{A}$ was defined by all the fully observed variables.
Another important decision when using PCA is the number of components to keep.
\cite{howardEtAl:2015} used only the first component in their simulations.
Since this component explained, on average, 40\% of the variance in the auxiliary data, they recommend using enough components to explain 40\% of the variance.
For our study, we generated more complex data for which a single component was not likely to suffice.
We, therefore, applied the intuitively appealing---albeit arbitrary---heuristic of using enough components to explain 50\% of the total variance in the data.

Running MI-QP in the high-dimensional procedure led to frequent convergence failures.
A more common use of the method includes accompanying the quickpred approach with a ridge penalty and data-driven checks that exclude collinear variables.
We decided to run MI-QP in this more favorable manner by applying the \textit{mice} package's usual data-screening procedures.
Accordingly, the \emph{mice()} call for MI-QP was specified with Bayesian normal linear regression as a univariate imputation method and with default values for the following arguments: ridge = $1e-5$, eps = $1e-4$, threshold = $0.999$.
Finally, we implemented the MI-AM method by applying the \emph{mice::mice()} function to only the analysis model variables with Bayesian normal linear regression as a univariate imputation method.
In this simulation study, the analysis model variables are the variables with missing values for which we wanted to estimate the means, variances, and covariances.

\subsubsection{Analysis and comparison criteria}\label{criteria}

The \emph{analysis model} comprised the joint distribution of the six variables with missing values.
Therefore, we refer to these six incomplete variables as the \emph{analysis model variables} below.
After imputation, we estimated the 6 means, 6 variances, and 15 covariances for these variables on each imputed data set and pooled the estimates via the \citet{rubin:1987} pooling rules.
We then compared the performances of the imputation methods by computing the bias, confidence interval coverage, and confidence interval width for each estimated parameter.

Since we generated multivariate normal data, the sample means, variances, and covariances were the sufficient statistics for the joint distribution of the analysis model variables.
Hence, we can infer that a method which demonstrates good performance when estimating these statistics will perform equally well when estimating other parameters that describe the same joint distribution.
For example, the slopes, $\beta = \Sigma_X^{-1}\Sigma_{X,y}$, intercept, $\alpha = \mu_y - \mu^T_X \beta$, and residual variance, $\sigma^2_{\varepsilon} = \sigma^2_y - \beta^T \Sigma_X \beta$ of a general linear model can be defined directly in terms of these statistics.
Using only this mean vector and covariance matrix, we could also factor analyze these six variables~\citep[pp. 53--55]{bartholomewEtAl:2011} or estimate their structural relations via a structural equation model~\citep[pp. 104--106]{bollen:1989}.
Importantly, the inverse implication does not generally hold.
For example, in the special case noted above wherein CC can produce unbiased slope estimates, the estimated means, variances, and covariances of the underlying data could still be biased unless the data were MCAR.
By focusing our analysis on a general set of sufficient statistics, we dissociated our results from any specific statistical model or test and increased the generalizability of our findings.

For a given parameter of interest $\theta$, we used the
absolute percent relative bias (PRB) to quantify the estimation bias introduced by the imputation procedure:
\begin{equation} \label{eq:prb}
	\textit{PRB} = \displaystyle\left\lvert\ \frac{\bar{\hat{\theta}} - \theta}{\theta} \right\rvert \times 100,
\end{equation}
where $\theta$ is the true value of the focal parameter defined as $\sum_{s=1}^{S} \hat{\theta}_{s}^{GS}/S$, with $\hat{\theta}_{s}^{GS}$ being the Gold Standard parameter estimate for the $s$th repetition.
The averaged focal parameter estimate under a given missing data treatment was computed as
$\bar{\hat{\theta}} = \sum_{s=1}^{S} \hat{\theta}_{s}/S$, with $\hat{\theta}_{s}$ being the estimate obtained from the treated incomplete data in the $s$th repetition.
Following \cite{muthenEtAl:1987}, we considered $\text{PRB} > 10$ as indicative of problematic estimation bias.

To assess the performance in hypothesis testing and interval estimation, we evaluated the confidence interval coverage (CIC) of the true parameter value:
\begin{equation} \label{eq:cic}
	\textit{CIC} =  \frac{ \sum_{s=1}^{S} I(\theta \in \widehat{\textit{CI}}_s ) }{S},
\end{equation}
where $\widehat{\textit{CI}}_s$ is the confidence interval of the parameter estimate $\hat{\theta}_{s}$ in the $s$th repetition, and $I(.)$ is the indicator function that returns 1 if the argument is true and 0 otherwise.

CICs below 0.9 are usually considered problematic for 95\% CIs (\citealp[p. 52]{vanBuuren:2018}; \citealp[p. 340]{collinsEtAl:2001}) as they imply inflated Type I error rates.
High CICs (e.g., above 0.99) indicate CIs that are too wide, implying inflated Type II error rates.
Therefore, we considered CIs to show severe under-coverage (over-coverage) if CIC $< 0.9$ (CIC $> 0.99$).
From a testing perspective, a CIC can be considered as significantly different from the nominal coverage rate if the
magnitude of its difference from the nominal coverage proportion ($p_0$) is more than two times the standard error
of $p_0$, $\textit{SE}(p_0) = \sqrt{p_0 (1-p_0)/S}$ \citep{burtonEtAl:2006}.
In our simulation study, the nominal coverage probability was 95\%.
Therefore, we considered 95\% CI coverages outside the interval $[0.94, 0.96]$ to be significantly different from the nominal coverage rate.
We assumed normal sampling distributions for variances and covariances when computing and pooling their CIs.
This assumption is plausible under large sample conditions.

We also reported the average width of the confidence intervals (CIW), an indicator of statistical efficiency.
An imputation method with a narrower confidence interval indicates higher efficiency and is therefore preferable.
Nevertheless, the narrower CIW should not come at the expense of a lower than nominal CIC \citep[p. 52]{vanBuuren:2018}.

\subsection{Results}\label{sec:results}

\begin{figure}
	\centering
	\includegraphics{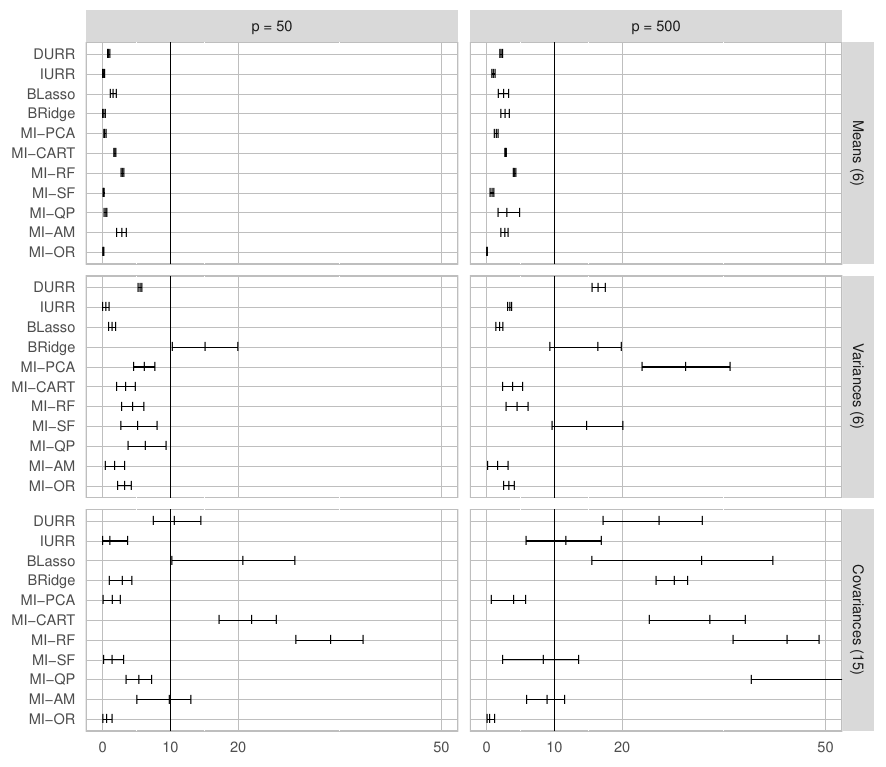}
	\caption{\label{fig:exp1bias}
		Minimum, average, and maximum absolute percent relative bias ($\text{PRB}$) for the 6 item means, 6 variances,
		and 15 covariances in the simulation study.
		If no data points are reported for a method in a panel, all of its PRBs were larger than 50.
		The methods reported on the Y-axis are: Direct Use of Regularized Regression (DURR), Indirect Use of Regularized
		Regression (IURR), MICE with Bayesian Lasso (BLasso), MICE with Bayesian Ridge (BRidge), MICE with Principal Component
		Analysis (MI-PCA), MICE with CART (MI-CART), MICE with Random Forests (MI-RF), MICE with step-forward selection (MI-SF), Oracle low-dimensional MICE (MI-OR), MICE with quickpred (MI-QP), MICE with Analysis Model (MI-AM), and Complete Case Analysis (CC).
	}
\end{figure}

\begin{figure}
	\centering
	\includegraphics{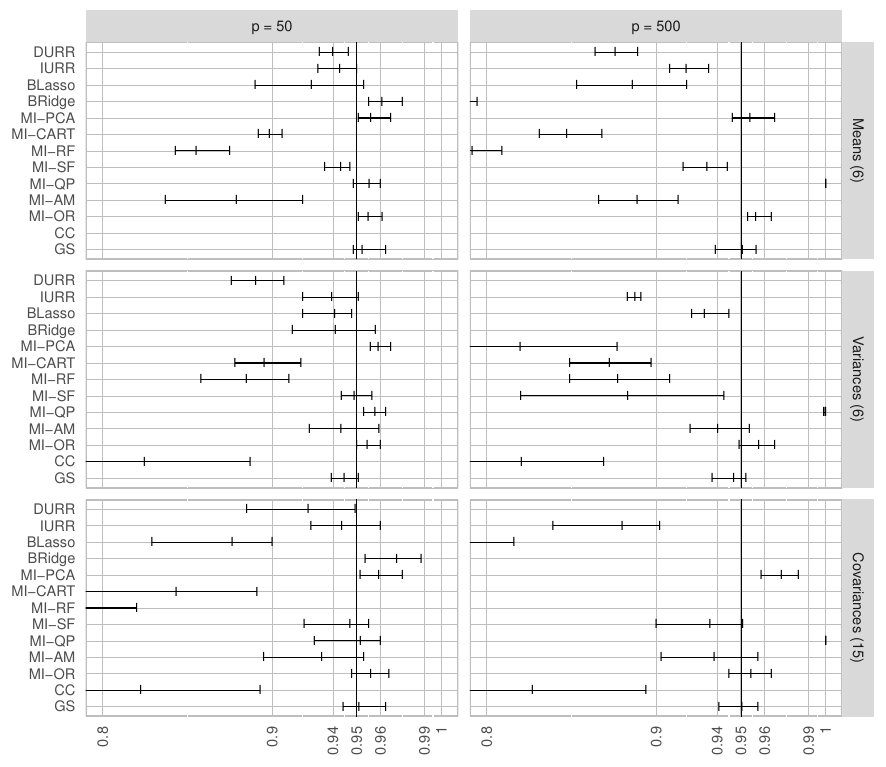}
	\caption{\label{fig:exp1cir}
		Minimum, average, and maximum confidence interval coverage (CIC) for the 6 item means, 6 variances,
		and 15 covariances in the simulation study.
		If no data points are reported for a method in a panel, all of its CICs were smaller than 0.80.
		The methods reported on the Y-axis are: Direct Use of Regularized Regression (DURR), Indirect Use of Regularized
		Regression (IURR), MICE with Bayesian Lasso (BLasso), MICE with Bayesian Ridge (BRidge), MICE with Principal Component
		Analysis (MI-PCA), MICE with CART (MI-CART), MICE with Random Forests (MI-RF), Oracle low-dimensional MICE (MI-OR), MICE with step-forward selection (MI-SF), MICE with quickpred (MI-QP), MICE with Analysis Model (MI-AM), Complete Case Analysis (CC), and Gold Standard Analysis (GS).
	}
\end{figure}

We computed both PRB and CIC for each of the 27 parameters in the analysis model (six means, six variances, and 15 covariances).
To summarize the results, we focus on the expected and extreme values of these measures.
In Figures \ref{fig:exp1bias} and \ref{fig:exp1cir}, we report the average, minimum, and maximum PRB and CIC obtained with the different missing data treatments, for each parameter type.
As the GS estimates were used to define the ``true'' values of the parameters, the bias for this method was by definition 0.
So, we do not include bias of the GS estimates in the figure.
For ease of presentation, we report the results only for the large proportion of missing cases ($pm = 0.3$) condition.
While the relative performances were independent of the missing data rate, the performance patterns were clearer with a larger proportion of missing values.
In the supplementary material, we included the same figures for the low proportion of missing cases ($pm = 0.1$) condition.
This article is accompanied by an interactive dashboard that is packaged as an R Shiny app~\citep{costantini:2023c}.
We recommend using this tool while reading the results and discussion sections to further elucidate the patterns of results discussed below.

\subsubsection{Means}

The largest $\text{PRB}$ for the means was below 10 for all imputation methods.
Only CC produced problematic degrees of bias.
Looking at the relative performances, IURR, BRidge, MI-PCA, MI-SF, and MI-OR resulted in smaller biases than the other methods.
In terms of CIC, only MI-PCA and MI-OR showed a consistently strong performance.
Neither method demonstrated any extreme under-/over-coverage (i.e., all CICs $\in$ [0.9, 0.99]), and both methods resulted in only the highest coverage being significantly different from nominal coverage (max CIC $> 0.96$).

IURR resulted in significant under-coverage of the true means (CICs $< 0.94$) in both the high-dimensional ($p = 500$) and low-dimensional ($p = 50$) conditions, although under-coverage was never severe (with CICs $\in [0.90, 0.94]$).
MI-SF resulted in similarly trivial under-coverage and always returned CICs $\in [0.90, 0.94]$.
DURR and BLasso demonstrated some significant differences from nominal coverage in the low-dimensional condition, and both led to extreme under-coverage in the high-dimensional condition.
The tree-based methods and CC performed most poorly.
These methods led to CICs significantly different from nominal coverage rates in all conditions, and they demonstrated extreme under-coverage even in the low-dimensional condition.
MI-QP resulted in close to nominal coverage in the low-dimensional condition, performing about as well as MI-OR and MI-PCA\@.
However, in virtually all replications of the high-dimensional condition, the CIs contained the true parameter values, thereby producing severe over-coverage.
Finally, MI-AM resulted in significant to extreme under-coverage.
This method was not influenced by the dimensionality of the data as it used the same six variables as predictors in both conditions.

\subsubsection{Variances}

IURR, BLasso, and the tree-based MI methods resulted in low biases (i.e., PRB $< 10$) in both the high and low dimensional conditions.
For BLasso, these low biases were paired with low deviations from nominal coverage rates.
IURR only demonstrated problematic CICs for the high-dimensional condition, where it produced extreme under-coverage (largest CIC $< 0.9$).
MI-CART and MI-RF did not produce reasonable coverage in either the low- or the high-dimensional condition, with the largest coverage being significantly different from nominal (CICs $< 0.94$) and the smallest being severely below the nominal level (CICs $< 0.9$).

MI-PCA and MI-SF showed acceptable biases and reasonable coverage rates in the low-dimensional condition, but they showed large biases and under-coverage in the high-dimensional condition.
In the low-dimensional condition, DURR produced low bias and reasonable coverage (i.e., only the lowest coverage being significantly different from nominal), but it resulted in PRBs $> 10$ for all variances in the high-dimensional condition, where it also produced extreme CI under-coverage.
BRidge and CC performed poorly in nearly all conditions.
These methods tended to demonstrate substantial biases and extreme under-coverage.
Although MI-QP performed well in the low-dimensional condition, in the high-dimensional condition, it resulted in PRBs larger than 50 and CICs close to 1 for all six item variances.
MI-AM maintained low bias and acceptable coverage for all item variances.

\subsubsection{Covariances}

MI-PCA was the only method that showed consistently strong performance when estimating covariances.
MI-PCA showed negligible bias and minimal deviations from nominal coverage in both low- and high-dimensional conditions.
In particular, the PRB was smaller than 10 for all covariances in both conditions and was almost as low as the PRB obtained by MI-OR\@.
MI-PCA never produced extreme under-/over-coverage, and when the CIC was significantly different from the nominal rate,
the CIs showed mild \emph{over}-coverage (i.e., CICs greater than 0.96 but smaller than 0.99).
After MI-PCA, IURR and MI-SF demonstrated the next strongest performance, with negligible bias and acceptable coverage in the low-dimensional condition.
However, in the high-dimensional condition, IURR produced large biases and extreme under-coverage with the average bias being above the 10\% threshold and even the largest coverage being just around the 90\% threshold.
In the high-dimensional condition, MI-SF showed a similar, albeit less severe, deterioration in performance.

Both MI-QP and BRidge displayed low bias and acceptable coverage in the low-dimensional condition, but they resulted in unacceptable biases in the high-dimensional condition.
In the high-dimensional condition, MI-QP led to 100\% coverage of the true values, while BRidge led to extreme under-coverage of the true values.
MI-AM, DURR, BLasso, and the tree-based MI methods tended to result in PRBs larger than 10, accompanied by under-coverage of the true covariance values, even in the low-dimensional conditions.

\subsubsection{Confidence interval width}

\begin{figure}
	\centering
	\includegraphics{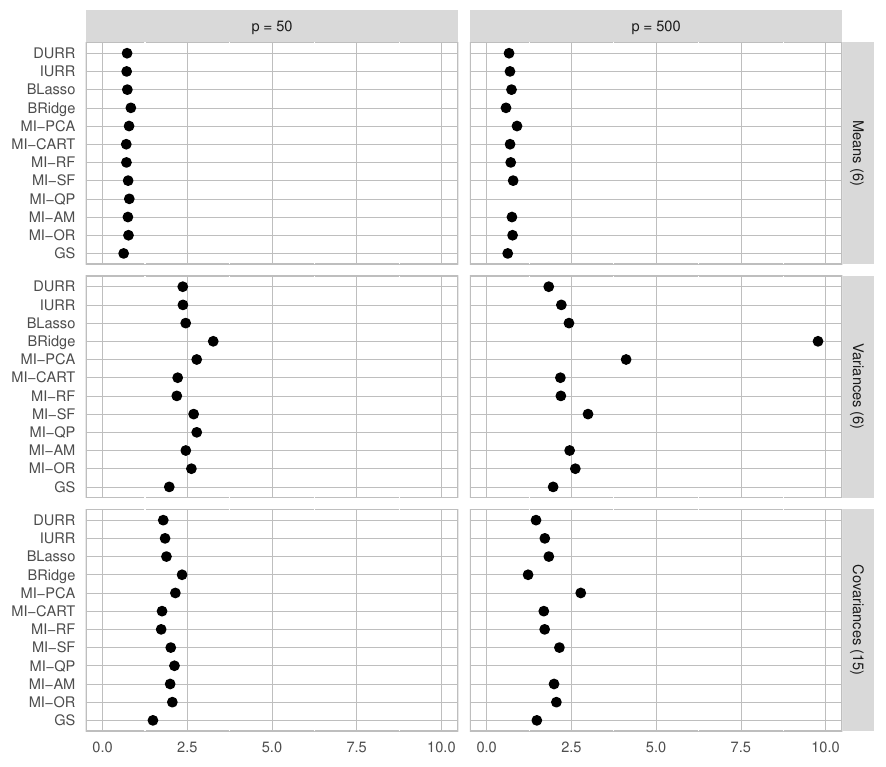}
	\caption{\label{fig:exp1ciw}
		Average confidence interval width (CIW) across the 6 item means, 6 variances,
		and 15 covariances in the simulation study.
		If no data points are reported for a method in a panel, its CIW was larger than 10.
		The methods reported on the Y-axis are: Direct Use of Regularized Regression (DURR), Indirect Use of Regularized
		Regression (IURR), MICE with Bayesian Lasso (BLasso), MICE with Bayesian Ridge (BRidge), MICE with Principal Component
		Analysis (MI-PCA), MICE with CART (MI-CART), MICE with Random Forests (MI-RF), MICE with step-forward selection (MI-SF), MICE with quickpred (MI-QP), MICE with Analysis Model (MI-AM), Oracle low-dimensional MICE (MI-OR), and Complete Case analysis (CC).
	}
\end{figure}

In Figure \ref{fig:exp1ciw}, we report the CIW obtained with the different missing data treatments, averaged per parameter type across the repetitions.
All methods maintained similar CIW independent of the dimensionality of the data.
The two exceptions were MI-QP and Bridge.
While the average CIW for MI-QP in the low-dimensional condition was in line with that of all the other methods, the CIW obtained with this method for all parameter types became larger than 10 for $P = 500$.
In the high-dimensional case, the item variance CIWs obtained by Bridge were four times as large as those obtained in the low-dimensional scenario.

\subsection{A note on collinearity}

Following feedback provided by a reviewer of an earlier draft of this article, we included an additional simulation study to explore the effect of collinearity.
We used the same simulation procedure described above, but we adjusted some of the design parameters.
We fixed the proportion of missing cases to the highest value ($pm = 0.3$), as this factor did not affect the relative performances of the methods.
We varied the number of columns in the data ($p \in \{50, 500\}$) and the strength of the correlation between the potential auxiliary variables ($\rho_{pav} \in \{0, 0.6, 0.8, 0.9\}$).
Correlations higher than 0.6 are unlikely in survey data, but including the higher levels provides the opportunity to explore how the imputation methods perform when faced with problematic levels of collinearity.

\begin{figure}
	\centering
	\includegraphics{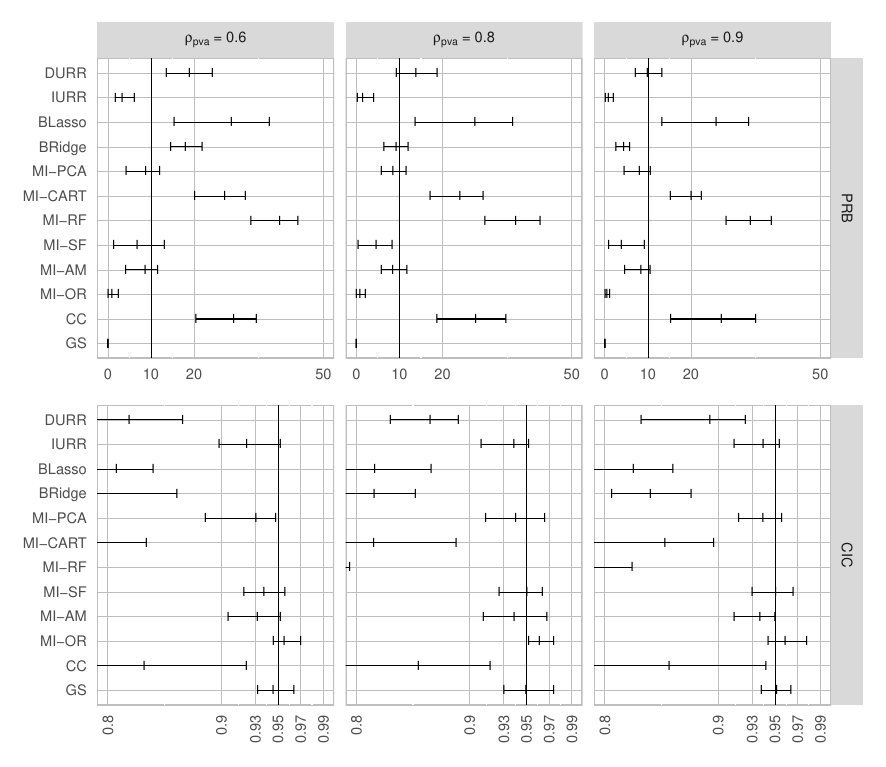}
	\caption{\label{fig:exp12}
		Minimum, average, and maximum absolute percent relative bias ($\text{PRB}$) confidence interval coverage ($\text{CIC}$) across 15 covariances estimated between items with imputed values.
		The methods reported on the Y-axis are: Direct Use of Regularized Regression (DURR), Indirect Use of Regularized
		Regression (IURR), MICE with Bayesian Lasso (BLasso), MICE with Bayesian Ridge (BRidge), MICE with Principal Component
		Analysis (MI-PCA), MICE with CART (MI-CART), MICE with Random Forests (MI-RF), MICE with step-forward selection (MI-SF), Oracle low-dimensional MICE (MI-OR), MICE with quickpred (MI-QP), MICE with Analysis Model (MI-AM), and Complete Case Analysis (CC).
	}
\end{figure}

In Figure~\ref{fig:exp12}, we report the average, minimum, and maximum PRB and CIC of the 15 covariances between two imputed items.
In this report, we focus on the high-dimensional condition ($p = 500$), and we omit $\rho_{pav} = 0$, which can be considered equivalent to the results already reported.
The interactive dashboard~\citep{costantini:2023c} contains the complete set of results.
The relative performances of the methods were mostly unchanged.
However, a few key differences should be noted.
First, shrinkage-based methods resulted in lower PRB and closer-to-nominal CIC for higher levels of $\rho_{pav}$.
In particular, the high PRB and low CIC that characterized BRidge in the original study results were mitigated as $\rho_{pav}$ increased.
For $\rho_{pav} = 0.9$, the highest bias returned by BRidge was lower than $10$, and the lowest CIC was higher than $0.80$.
Similar trends arose for IURR and DURR, for which higher values of $\rho_{pav}$ led to a lower PRB and closer-to-nominal CIC.
Second, for higher values of $\rho_{pav}$, the PRB and CIC from MI-PCA essentially mirrored those of MI-AM.
Finally, in the high-dimensional condition ($p = 500$), MI-QP had a prohibitively long imputation time.
In a small trial run of the simulation, MI-QP required around 360 minutes to impute a single data set generated with $\rho_{pav} = 0.6$ and around 1130 minutes to impute a data set generated with $\rho_{pav} = 0.9$.
IURR and MI-SF, the next two most computationally intensive methods, each took around 10 minutes to impute these data sets.
Consequently, we included MI-QP only in the low dimensional condition of this additional simulation study.

    \section{EVS resampling study}

	We performed a resampling study based on the EVS data to assess whether the results of our simulation study would replicate in more realistic data.
	EVS is a high-quality survey widely used by sociologists for comparative studies between European countries \citep{EVSbib}.
	Furthermore, it is freely available and represents the type of data social scientists regularly analyze.
	Variables in the EVS data are discrete numerical and categorical items following a variety of distributions.

	To perform the resampling study, we treated the original EVS data as a population.
	We then resampled $S = 1000$ data sets of $n$ units from this population, and we used these replicates as we used the multivariate normal samples in the
	simulation study.
	For each replicate, we imposed missing values, and we treated these missing values with the same methods explored in the simulation study.
	This procedure was repeated for low-dimensional and high-dimensional conditions.
	As the number of predictors in the data was fixed at $p = 243$, we controlled the dimensionality of the data by varying the sample size ($n \in \{1000, 300\}$).
	When the sample size was 300, after dummy coding categorical predictors, even a small proportion of missing values ($pm = 0.1$) led to a high-dimensional ($p > n$) situation.
	Although $n = 300$ might be too low to represent the typical use of EVS data, we do not see this as a limitation on the following results, for two reasons.
	First, our purpose in conducting this resampling study was primarily to see if our simulation results would carry over into data generated from a more realistic population model, not necessarily to see if those results would hold in a typical social science data set.
	Increasing the sample size to match ranges typically seen in analyzes of EVS data would remove the high-dimensional condition where we saw the most interesting results in the simulation, thereby greatly reducing the utility of the resampling study.
	Second, many social science studies analyze data with around 300 observations, so our samples are not unrealistic in a general sense.

\subsection{Resampling study procedure} \label{resProc}

\subsubsection{Data preparation and sampling}

	We used the third prerelease of the 2017 wave of EVS data \citep{EVS:2017} to create a population data set with \emph{no missing values}.
	The original data set contained 55,000 observations from 34 countries.
	We selected only the four founding countries of the European Union included in the data set (France, Germany, Italy, and the Netherlands)
	because keeping all countries would have entailed either including a set of 33 dummy codes in the imputation models or imputing under some form of a multilevel model.
	Since both of these options fall outside the scope of the current study, we opted to subset the data as described.
	We excluded all columns that contained duplicated information (e.g., recoded versions of other variables), or metadata (e.g., time of the interview, mode of data collection).

	The original EVS data set contained missing values.
	We needed to treat these missing data before we could use the EVS data in the resampling study.
	We used the \emph{mice} package to fill the missing values with a single round of predictive mean matching (PMM).
	We used the \emph{quickpred} function, to select the predictors for the imputation models.
	We implemented the variable selection by setting the minimum correlation threshold in \emph{quickpred} to 0.3.
	The number of iterations in the \emph{mice()} run was set to 200.
	We used a single imputation, and not MI, because this imputation procedure was used only to obtain a set of pseudo-fully observed data to act as the population in our resampling study and not for statistical modeling, estimation, or inference with respect to the true population from which the EVS data were sampled.
	For the same reason, the relatively poor performance that we observed for MI-QP in the simulation study is not relevant here.
	At the end of the data cleaning process, we obtained a pseudo-fully observed data set of 8,045 observations across four countries with $p = 243$ variables.
	For every replicate in the resampling study, we generated a bootstrap sample by sampling $n$ observations with replacement from this data set.

\subsubsection{Analysis models}

	To define plausible analysis models, we reviewed the models reported in the repository of publications using EVS data that is available on the EVS website \citep{EVSbib}.
	As a result, we defined two linear regression models.
	Model 1 was inspired by \cite{koneke:2014}.
	The dependent variable was a 10-point item measuring euthanasia acceptance (`Can [euthanasia] always be justified, never be justified, or something in between?').
	The predictors included an item measuring the self-reported importance of religion in one's life, trust in the health care system, trust in the state, trust in the press, country, sex, age, education, and religious denomination.
	A researcher might estimate this model to test a hypothesis regarding the effect of religiosity on the acceptance of end-of-life treatments.

	Model 2 was inspired by \cite{immerzeel:2015}.
	The dependent variable was a harmonized variable that quantifies the respondents' tendencies to vote for left- or right-wing parties, expressed on a 10-point left-to-right continuum.
	The predictors included a scale measuring respondents' attitudes toward immigrants and immigration (`nativist attitudes scale').
	The scale was obtained by taking the average of respondents' agreement, on a scale from 1 to 10, with three
	statements: `immigrants take jobs away from natives', `immigrants increase crime problems', and
	`immigrants are a strain on welfare system'.
	The remaining predictors were:
	attitudes toward law and order, attitudes toward authoritarianism, interest in politics, level of political activity,
	country, sex, age, education, employment status, socioeconomic status, importance of religion in life, religious denomination, and the size of the town where the interview was conducted.
	A researcher might estimate this model to test a hypothesis regarding the effect of xenophobia on voting tendencies.

\subsubsection{Missing data imposition}

	We imposed missing data on six variables using the same strategy as in the simulation study.
	The targets of missing data imposition were the two dependent variables in models 1 and 2 (i.e., euthanasia acceptance, and left-to-right voting tendency), religiosity, and the three items making up the ``nativist attitudes'' scale.
	The response model was the same as in Equation \eqref{eq:rm}, and three variables were included in $\tilde{\bm{Z}}$: age, education, and an item measuring trust in new people\footnote{
		The pseudo R-squared for the logistic regression of the missing value indicator on the predictors of missingness was around 14\%.
		The AUC for the logistic regression was around 0.75.
	}.
	We chose these predictors because older people tend to have higher item nonresponse rates than younger people, and lower educated people tend to have higher item non-response rates than higher educated people \citep{guadagnoliCleary:1992, leeuwEtAl:2003}.
	We also assumed that people with less trust in strangers would have a higher nonresponse tendency as they are likely to withhold more information from the interviewer (a stranger).

\subsubsection{Imputation}

	We treated the missing values with the same methods used in the simulation study.
	MI-AM used all the variables present in either of the analysis models as predictors for the imputation models.
	MI-PCA was performed considering all the fully observed variables as possible auxiliary variables.
	In other words, the six variables with missing values were used in their raw form, while the remaining 237 were used to extract PCs.
	The other imputation methods were parameterized in the same way as in the simulation study, and convergence checks were performed in the same way.
	These convergence checks suggested that the imputation models had converged after 60 iterations.

\subsection{Results}

\begin{figure}
	\centering
	\includegraphics[height=\linewidth, keepaspectratio]{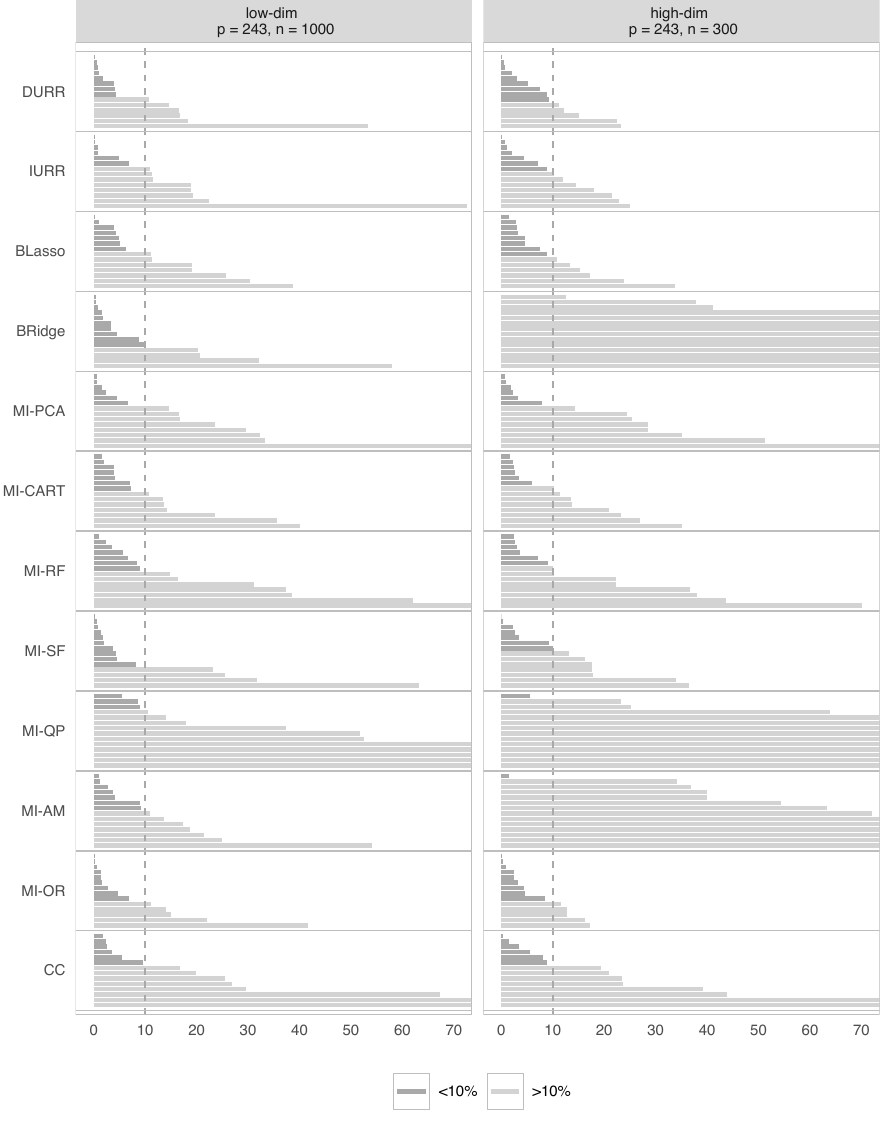}
	\caption{
		PRB for all the model parameters in model 2.
		For each method, the PRBs are ordered by increasing absolute value.
		The methods reported on the Y-axis are: Direct Use of Regularized Regression (DURR), Indirect Use of Regularized
		Regression (IURR), MICE with Bayesian Lasso (BLasso), MICE with Bayesian Ridge (BRidge), MICE with Principal Component
		Analysis (MI-PCA), MICE with CART (MI-CART), MICE with Random Forests (MI-RF), MICE with step-forward selection (MI-SF), Oracle low-dimensional MICE (MI-OR),
		MICE with quickpred (MI-QP), MICE with analysis model (MI-AM), and Complete Case analysis
		(CC).
	}
		\label{fig:exp4_bias_allP}
\end{figure}

\begin{figure}
	\centering
	\includegraphics[height=\linewidth, keepaspectratio]{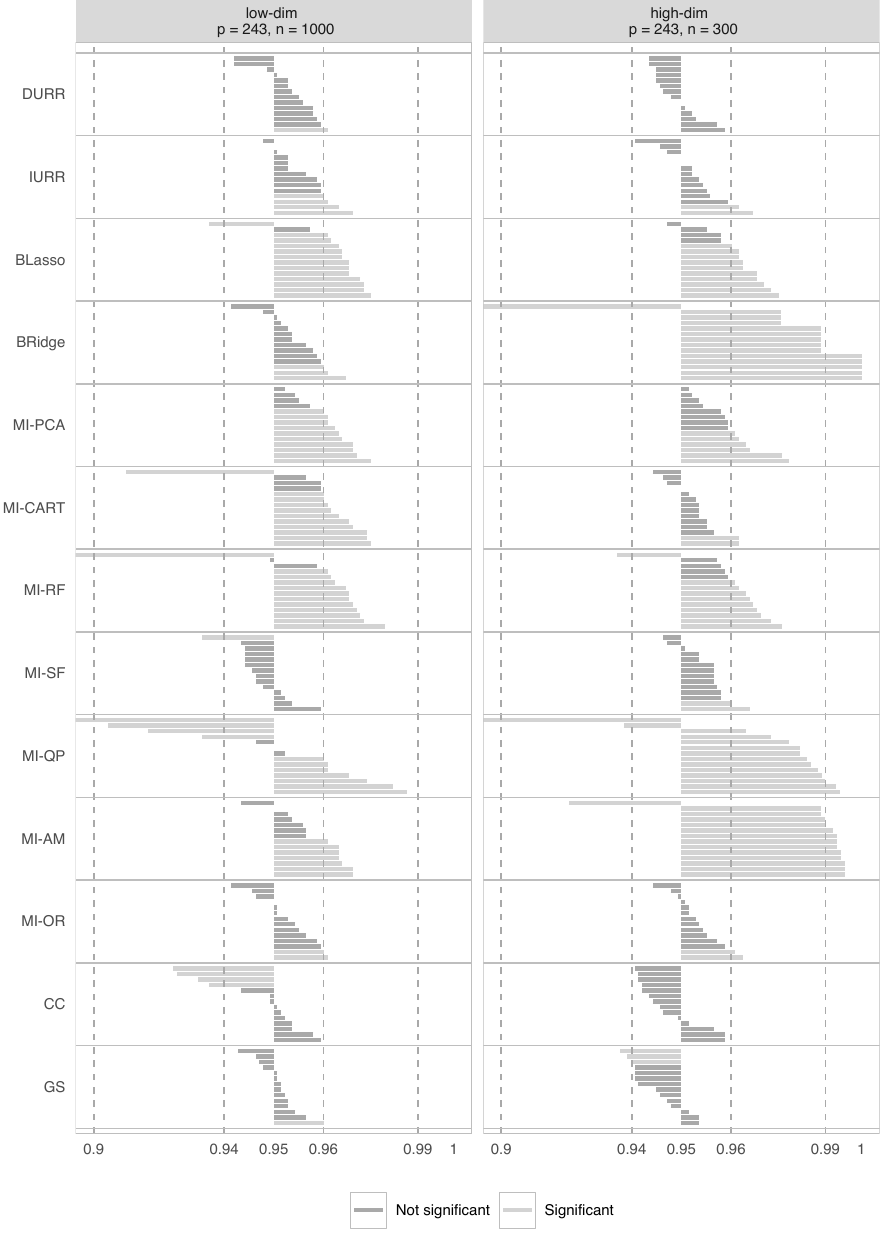}
	\caption{
		CIC for all model parameters in model 2.
		For each method, the CICs are ordered by increasing value.
		The methods reported on the Y-axis are: Direct Use of Regularized Regression (DURR), Indirect Use of Regularized
		Regression (IURR), MICE with Bayesian Lasso (BLasso), MICE with Bayesian Ridge (BRidge), MICE with Principal Component
		Analysis (MI-PCA), MICE with CART (MI-CART), MICE with Random Forests (MI-RF), MICE with step-forward selection (MI-SF), MICE with quickpred (MI-QP), MICE with analysis model (MI-AM), Oracle low-dimensional MICE (MI-OR), Complete Case analysis (CC), and Gold Standard analysis (GS).
	}
	\label{fig:exp4_ci_allP}
\end{figure}

\begin{figure}
	\centering
	\includegraphics[height=\linewidth, keepaspectratio]{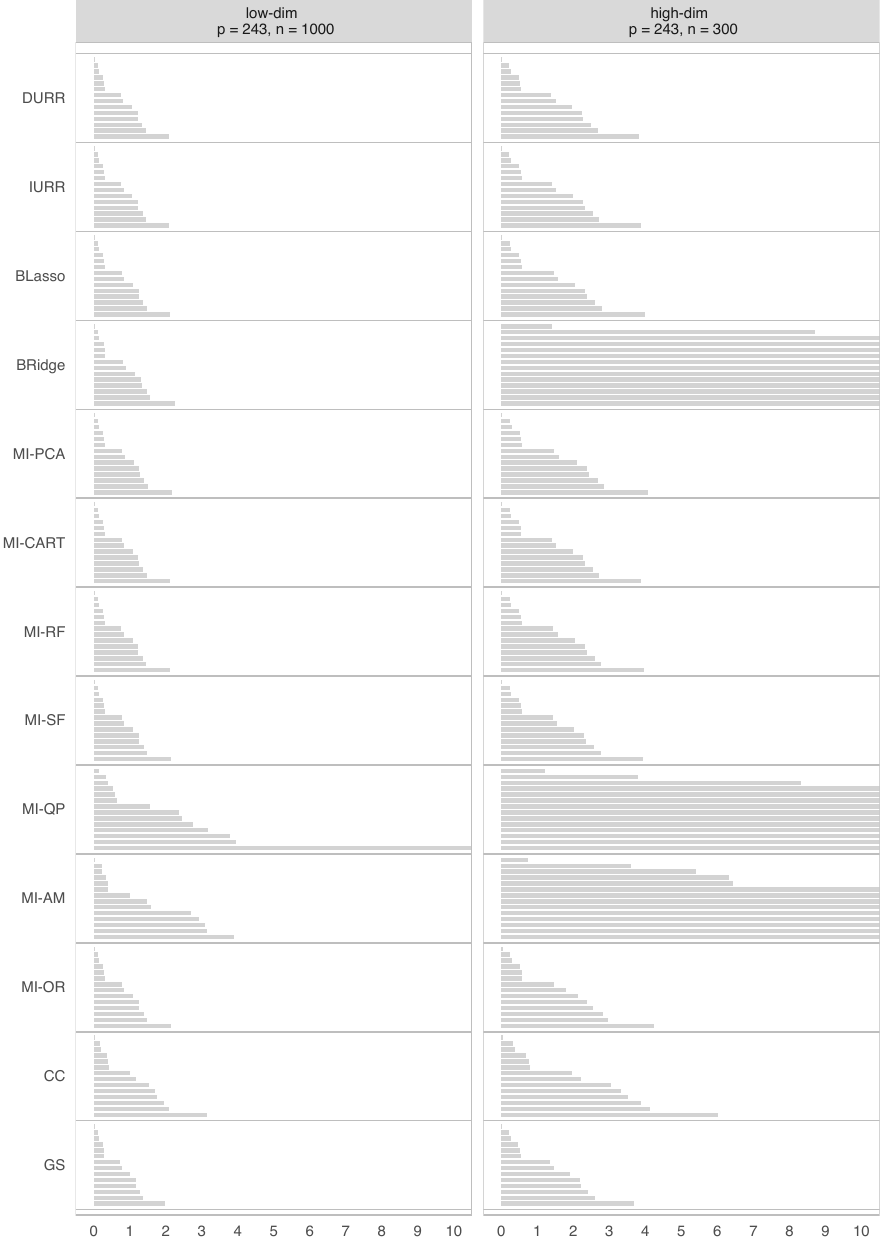}
	\caption{
		Average CIW for all model parameters in model 2.
		For each method, the CICs are ordered by increasing value.
		The methods reported on the Y-axis are: Direct Use of Regularized Regression (DURR), Indirect Use of Regularized
		Regression (IURR), MICE with Bayesian Lasso (BLasso), MICE with Bayesian Ridge (BRidge), MICE with Principal Component
		Analysis (MI-PCA), MICE with CART (MI-CART), MICE with Random Forests (MI-RF), MICE with step-forward selection (MI-SF), MICE with quickpred (MI-QP), MICE with analysis model (MI-AM), Oracle low-dimensional MICE (MI-OR), Complete Case analysis (CC), and Gold Standard analysis (GS).
	}
	\label{fig:exp4_ciw_allP}
\end{figure}

	When estimating linear regression models, all partial regression coefficients can be influenced by missing values on a subset of the variables included in the model.
	Therefore, it is important to evaluate the estimation bias and CIC rates for all model parameters.
	Figure \ref{fig:exp4_bias_allP} reports the absolute PRBs for the intercept and all partial slopes from Model 2 obtained after using each imputation method, for both the low- and high-dimensional conditions.
	Model 2 has an intercept and 13 regression coefficients.
	Every horizontal line in the figure represents the PRB for the estimation of one of these 14 parameters.
	Figures \ref{fig:exp4_ci_allP} and \ref{fig:exp4_ciw_allP} report CIC and CIW results in the same way.
	For ease of presentation, results for Model 1 are reported in the supplementary materials.

	As seen in Figure \ref{fig:exp4_bias_allP}, in both the high- and low-dimensional conditions, DURR, IURR, BLasso, MI-CART, and MI-SF showed only slightly larger PRBs than MI-OR\@.
	However, even MI-OR did not provide entirely unbiased parameter estimates.
	After imputing with MI-OR, almost half of the parameters in Model 2 were estimated with large bias (PRB $> 10\%$).
	MI-PCA, MI-RF, and CC showed similar trends but produced larger PRBs (particularly CC).
	BRidge demonstrated the same results described in the simulation studies.
	It was competitive in the low-dimensional scenario, but it was inadequate with high-dimensional data (all PRBs $> 10$.)
	In the low-dimensional condition, MI-QP resulted in only three parameter estimates with acceptable bias and only one in the high-dimensional condition.
	MI-AM resulted in six parameter estimates with acceptable bias in the low-dimensional condition but only one in the high-dimensional condition.

	As shown in Figure \ref{fig:exp4_ci_allP}, MI-SF, MI-OR, and DURR resulted in the lowest deviations from nominal coverage, with only one or two coverages differing significantly from the nominal level.
	IURR showed a similar trend but four coverages were significantly different from nominal in the low-dimensional condition.

	BLasso, MI-PCA, MI-CART, MI-RF, MI-SF, and MI-AM all showed similar performance in the low-dimensional condition.
	These methods all significantly over-covered most of the parameters but did not produce any extreme under-/over-coverage, except for one parameter for MI-RF.
	BLasso, MI-PCA, and MI-RF maintained similar performance in the high-dimensional condition, but MI-CART improved to match the performance of MI-OR, and MI-AM produced extreme over-coverage for most of the parameters.
	BRidge performed well in the low-dimensional condition---around the level of IURR---but produced very poor coverages in the high-dimensional condition.
	MI-QP performed poorly in both the low- and high-dimensional conditions, producing only two non-significant coverages in the low-dimensional condition and none in the high-dimensional condition.
	CC performed quite well, but it had a much more pronounced tendency toward \emph{under}-coverage than the MI methods.
	Notably, very few of the CICs fell into the range of extreme under-/over-coverage.
	Only the high-dimensional estimates from BRidge and MI-AM consistently exhibited extreme under-/over-coverage.

	Finally, the average CIW for every parameter estimate is reported in Figure \ref{fig:exp4_ciw_allP}.
	In the low-dimensional condition, all methods result in similar CIWs.
	All methods result in larger confidence intervals in the high-dimensional condition reflecting a natural loss of information due to the smaller sample size used.
	However, Bridge, MI-QP, and MI-AM show drastically larger CIWs for the majority of the parameters.

\subsubsection{Imputation time}

\begin{figure}
	\centering
	\includegraphics{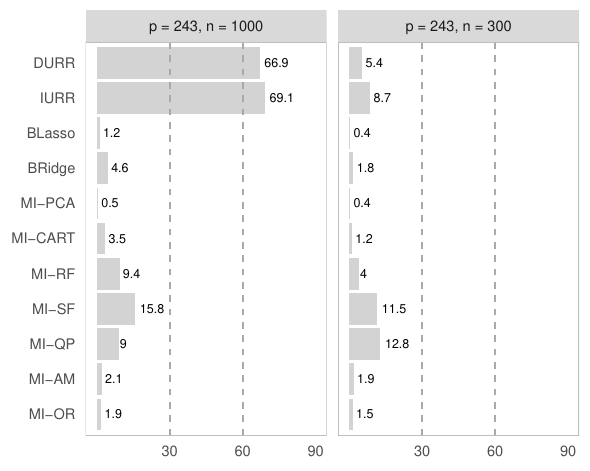}
	\caption{
		Average imputation time in minutes for the different MI methods when applied to the two different resampling study conditions.
	}
	\label{fig:exp4_time}
\end{figure}

	Figure \ref{fig:exp4_time} reports the average imputation time for the different methods.
	IURR and DURR were the most time-consuming methods, with imputation times above one hour in the low-dimensional condition.
	In the high-dimensional condition, IURR and DURR were not as time-intensive due to the smaller sample size but still took more than ten times longer than MI-PCA and BLasso.
	MI-PCA was the fastest method, with imputation times of under a minute in both the high- and low-dimensional conditions.
	BLasso, MI-OR, and MI-AM were close seconds, with imputation times of two minutes or less in both conditions.
	BRidge, MI-CART, MI-RF, MI-SF, and MI-QP fell in the middle, with imputations times ranging from 3.5 (MI-CART) to 15.8 (MI-SF) minutes in the low-dimensional condition and from 1.2 (MI-CART) to 12.8 (MI-QP) minutes in the high-dimensional condition.

    \section{Discussion}

\subsection{Methods that work well}

	On balance, IURR, MI-SF, and MI-PCA were the strongest performers across the simulation study and the resampling study.
	In the simulation study, IURR and MI-SF produced trivial estimation bias for all parameters in the low-dimensional condition and for the means in the high-dimensional condition.
	Furthermore, the covariance estimation bias introduced by these two methods in the high-dimensional condition only slightly exceeded the PRB $= 10$ threshold, while most of the other MI methods resulted in covariance PRBs larger than 20 (with MI-PCA being the most salient exception).
	IURR and MI-SF produced good coverages in the low-dimensional condition but tended to under-cover in the high-dimensional condition, especially for variances and covariances.
	In the resampling study, IURR and MI-SF were also among the strongest performers.
	Although they did not demonstrate the best performance, there were no conditions in which IURR or MI-SF produced unacceptable results.

	The confidence interval widths of IURR and MI-SF were in line with that of the other methods.
	In the simulation study, the confidence intervals produced by these methods were not influenced by the dimensionality of the data.
	In the resampling study, their confidence intervals were wider in the high-dimensional condition than in the low-dimensional one.
	However, this was the same pattern that affected most methods and it was caused by the smaller sample size we used to achieve the $p > n$ scenario in a data set with a fixed number of predictors.
	Overall, the confidence interval width pattern followed by IURR and MI-SF suggests that their imputation precision is not affected by a larger number of possible predictors.

  From the end-user's perspective, IURR is an appealing method.
  IURR does not require the imputer to make choices regarding which variables are relevant for the imputation procedure.
	The only additional decision required of the imputer is selecting the number of folds to use when cross-validating the penalty parameter.
	As a result, an IURR imputation run is easy to specify, which makes IURR an appealing method for the imputation of large social scientific data sets.
	However, IURR is relatively computationally intensive.
	If the number of variables with missing values is large, IURR might result in prohibitive imputation time.

	Similarly, an MI-SF run is easy to specify and only requires the user to choose the minimum sufficient increase in $R^2$ to use in the step-froward algorithm.
	However, the lack of clear guidelines on how to tune this parameter introduces more researcher's degrees of freedom than other methods.
	Finally, the imputation time of MI-SF was among the longest of the methods we considered.

	In the simulation study, MI-PCA showed small bias and good coverage for both item means and covariances.
	Although it exhibited a large bias of the item variances, the---arguably more interesting---covariance relations between variables with missing values were always correctly estimated.
	Notably, MI-PCA was the only method resulting in small bias and close-to-nominal CIC for the covariances, even in the high-dimensional condition.
	When the CICs obtained with MI-PCA deviated significantly from nominal rates, they over-covered.
	In most situations, over-coverage is less worrisome than under-coverage as it leads to conservative, rather than liberal,
	inferential conclusions.
	In terms of confidence interval width, MI-PCA demonstrated the same pattern as IURR.
  In the resampling study, MI-PCA demonstrated middle-of-the-pack performance: somewhat worse than IURR, but still within acceptable levels.
	
	In the additional simulation study evaluating the effects of collinearity, MI-PCA resulted in the same bias and confidence interval coverage as MI-AM when the potential auxiliary variables were highly correlated.
	This trend was caused by a subtle interaction between the data-generating model and the rule used to select the number of PCs.
	In every condition, there were only four true MAR predictors out of the pool of either 44 or 494 potential auxiliary variables.
  Consequently, the manner in which these four MAR predictors were represented in the component scores played a crucial role in the performance of MI-PCA.
  When $\rho_{pav}$ was relatively small (i.e., the potential auxiliary variables were not strongly correlated), retaining enough components to explain 50\% of the variance tended to select approximately 20 PCs.
  Furthermore, the first of these components was predominately defined by the four MAR predictors, since these four variables comprised the entire subset of predictor data with non-trivial correlations.
  For high values of $\rho_{pav}$, however, the behavior of the MI-PCA algorithm shifted in two important ways.
  First, due to the increased homogeneity of the data, the first PC explained a much larger proportion of the total variance, so the 50\% rule selected only one PC.
	Second, the first PC was predominately defined by the noise variables, since their high associations represented the majority of the reliable variance in the data.
  As a result, for large values of $\rho_{pav}$, the imputation models used by MI-PCA differed from the MI-AM imputation models only by adding a principal component that primarily summarized the noise variables as another useless predictor.
	A detailed explanation of this phenomenon is presented in module 3 of the interactive dashboard \citep{costantini:2023c}.

  Importantly, this finding does not suggest that MI-PCA cannot treat highly collinear data. 
  Rather, the poor performance seen here suggests that heuristic decision rules---such as keeping the first PC or enough components to explain 50\% of the total variance---should not be mindlessly applied when running MI-PCA. 
	Using a different non-graphical decision rule (e.g., the Kaiser criterion, \citealp{guttman:1954}; \citealp{kaiser:1960}) should preclude the problem described above and allow MI-PCA to compete with other automatic model-building strategies.
	
  On balance, we believe the strong performance demonstrated by MI-PCA in the simulation study outweighs the mediocre performance shown in the resampling study.
	Furthermore, as noted above, the poor performance of MI-PCA in the high-collinearity study merely represents a weakness of our current implementation, not a general flaw in the underlying method.
  Consequently, we view MI-PCA as a promising approach for data analysts interested in testing theories on large social scientific data sets with missing values.
	
\subsection{Methods with mixed results}

	In both the simulation study and the resampling study, BRidge manifested the same mixed performance.
	This method worked well when the imputation task was low-dimensional but led to extreme bias and unacceptable CI coverage in nearly all the high-dimensional conditions.
	Furthermore, the high-dimensionality of the data led to much wider confidence intervals compared to the ones obtained by other methods.
	Our results suggest that BRidge is effective only for low-dimensional imputation problems or in the presence of highly collinear data.
	The poor performance of BRidge compared to the other shrinkage methods might be explained by the fact that BRidge used a fixed ridge penalty across all iterations, while DURR, IURR, and BLasso allowed the penalty parameter to adapt to the improved imputations.

	As implemented here, MI-QP was only effective in low-dimensional settings.
	The instability of MI-QP in high-dimensional scenarios was apparent not only because of its larger bias but also its very wide confidence intervals.
	The much wider confidence intervals obtained by MI-QP in the high-dimensional scenario resulted in a 100\% coverage for the 95\%-confidence intervals, despite the large bias, revealing grossly imprecise imputations.
	MI-QP is also unable to address collinearity, as it selects predictors based on their bivariate relations with the variable under imputation and its missing data indicator without considering associations between the selected predictors.
	Hence, when faced with many highly correlated predictors, MI-QP can also be extremely computationally intensive due to the need to invert near-singular matrices.

	DURR performed very well in the resampling study and quite poorly in the simulation study.
	In the resampling study, DURR was probably the best overall method in terms of bias and coverage, but it performed very badly in the high-dimensional condition of the simulation study.
	In the simulation study's low-dimensional condition, DURR produced small bias, good CI coverage, and similar CIW to IURR for item means and variances.
	However, compared to IURR, it suffered from greater deterioration in performance when applied to high-dimensional data, especially in terms of coverage.
	Our results suggest that DURR may have some unique benefits when treating the types of more discrete data seen in the resampling study.
	On balance, though, DURR probably should not be preferred to IURR\@.
                
	There was little difference in performance between the use of CART and random forests as elementary imputation methods within the MICE algorithm.
	In line with what \cite{dooveEtAl:2014} found, when a difference was noticeable, the simpler CART generally outperformed the more complex random forests.
	Both MI-CART and MI-RF produced large covariance bias in the simulation study.
	Although the bias for means and variances was acceptable, it was usually larger than that obtained by other MI methods.
	Furthermore, in terms of CI coverage, both methods showed a large under-coverage of the true values in the high-dimensional condition.
	In the resampling study, MI-CART and MI-RF both showed somewhat better performance than in the simulation study but not enough better to outweigh the mediocre simulation study performance.
	Although the nonparametric nature of these approaches elegantly avoids over-parameterization of imputation models, these methods were still outperformed by IURR and MI-PCA\@.

	In the simulation study, BLasso resulted in small biases for item means and variances, even in the high-dimensional conditions, but it produced unacceptably biased covariance in both the low- and high-dimensional conditions.
	On the other hand, BLasso seemed to recover the relationships between variables in the resampling study well, where the overall bias levels for the regression coefficients were similar to those of MI-OR\@.
	However, in terms of CI coverage, BLasso showed poor performance in both studies resulting in either under-coverage or over-coverage for most parameters in the high-dimensional conditions.

	The mixed performance of BLasso is also accompanied by a few obstacles to its application for social scientific research.
	Using \cite{hans:2010}'s Bayesian Lasso requires the specification of six hyper-parameters, which introduces more researcher degrees of freedom and demands a strong grasp of Bayesian statistics.
	Furthermore, the method has not currently been developed for multi-categorical data imputation, a common task in the social sciences.
	As a result, we do not recommend BLasso for the imputation of large social science data sets.

	Finally, we do not recommend using MI-AM to impute large social science data sets.
	MI-AM bypasses the need to select which of the many potential auxiliary variables should be included in the imputation models by using only the analysis model variables as predictors.
	Therefore, MI-AM can be effective if the MAR predictors are part of the analysis model, but, as shown in the simulation study, it can lead to biased parameter estimates if they are not.
	In our simulation study, smaller biases and better coverages could always be achieved by using at least one of the alternative methods we evaluated.

    \section{Limitations and future directions}

	The present study aimed to compare current implementations of existing imputation methods.
	As a result, the scope of the simulation and resampling studies was limited by the current development state of the different methods.
	For example, DURR, IURR, and MI-PCA allow imputation of any type of data:
	DURR and IURR have been developed for categorical data imputation \citep{dengEtAl:2016}, and MI-PCA can be performed with any standard imputation model for categorical data.
	However, BLasso has not been formally developed for imputing multi-categorical variables yet.
	This limitation of BLasso forced us to work with missing values on variables that are either continuous or usually considered as such in practice (e.g., Likert-type scales).
	To maintain a fair comparison with BLasso, all methods were implemented with the assumption that the imputed variables are continuous and normally distributed.
	However, IURR, DURR, and MI-PCA could have performed differently in the resampling study if we had used their ordinal data implementations.

	More generally, the results reported in this article only apply to the specific implementations of the algorithms we used.
	Many of the methods discussed could have been implemented differently.
	\cite{zhaoLong:2016} proposed versions of IURR and DURR using the elastic net penalty \citep{zouHastie:2005} and the adaptive lasso \citep{zou:2006} instead of the lasso penalty.
	Although no substantial performance differences between penalty specifications emerged from the work of \cite{zhaoLong:2016} or \cite{dengEtAl:2016}, we must acknowledge that we did not investigate
	the impact of different types of regularization in the present study.
	Similarly, we have not investigated the sensitivity of BLasso to different hyper-parameters choices.
	Furthermore, the use of random forests within the MICE algorithm followed \cite{dooveEtAl:2014}, the version supported in the popular R package \emph{mice}.
	However, \cite{shahEtAl:2014} independently developed another implementation of random forests within the MICE algorithm, which was available in the now archived R package \emph{CALIBERrfimpute} \citep{CALIBERrfimpute}.
	We are not aware of any evidence or theoretical reason to expect differences between the two implementations, but we did not verify this empirically.
	Finally, there are many alternatives to ordinary least square (OLS) estimation that we did not consider.
	\cite{dempsterEtAl:1977b} compared the properties of 57 such OLS alternatives, including different variants of ridge regression, subset regression (e.g., forward and backward model selection), and principal component regression, when applied to fully observed data.
	Any of these variants could be used as the elementary imputation model in a MICE implementation.
	In the present study, however, our inclusion criteria for imputation methods precluded consideration of these alternatives. 
	We considered only those high-dimensional prediction methods that have already been recommended in the literature specifically for MI.
	This is the same reason we did not consider many state-of-the-art prediction methods like (deep) neural networks or support vector machines/regressions, even though those methods currently dominate all others in terms of raw prediction and classification performance.
	
  Our implementation of MI-PCA was limited in several ways.
	First, MI-PCA requires choosing the number of components to extract from the auxiliary variables.
	In this study, we decided to retain the first components that explained 50\% of the total variance in the auxiliary variables.
	However, this decision was arbitrary, and the results of collinearity-focused simulation study clearly demonstrate some of the possible deleterious consequences of this approach.
	Additionally, the good performance of MI-PCA may have been partially driven by the fact that, while imputing the $j$th variable, all other variables under imputation were used directly as predictors.
	If the other variables under imputation had been included in the imputation models through the PCs extraction step, and not used as separate, individual predictors, the performance of MI-PCA might have been less favorable.
	In \cite{costantiniEtAl:2023} we assess the effects of these two factors on the MI-PCA method.
	The unsupervised nature of the classical PCA through which MI-PCA constructs imputation model predictors may also be a limiting feature.
	While classical PCA should optimally distill the variance of the potential auxiliary variables into a succinct set of component scores, these component scores may not be useful predictors in the imputation model (e.g., if most of the potential auxiliary variables were not good predictors to begin with).
	Supervised versions of PCA (e.g., supervised PCA, \cite{bairEtAl:2006}, principal covariates regression, \cite{deJongKiers:1992}) could overcome this limitation.
	In \cite{costantiniEtAl:2022c} we evaluate the performance of MI-PCA when the component scores are extracted via several different supervised versions of PCA.

    \section{Conclusions}
	Our objective in this project was to find a good data-driven way to select the predictors that go into an imputation model.
	A wide range of methods have been proposed to address this issue, but little research has been done to compare their performance.
	With this article, we start to fill this gap and provide initial insights into applying such methods in social science research.
	IURR, MI-SF, and MI-PCA showed promising performance when compared to other high-dimensional imputation approaches.
	While all of these methods represent good options for automatically defining the imputation models of an MI procedure, MI-PCA is the more practically appealing option due to its much greater speed.
	However, the current implementation of MI-PCA is limited, and making the most of this method will require further research and optimization, especially regarding methods for the number of components.
	Finally, Bayesian ridge regression is a good alternative when the imputer wants to have an automatic way of defining the imputation models in a low-dimensional setting ($n \gg p$).

    \setcounter{secnumdepth}{4}

    \section{Disclosure statement}
    No authors reported any financial or other conflicts of interest in relation to the work described.

    \section{Funding} 
    No funds, grants, or other support were received.

    \section{Author's Note}
    Edoardo Costantini is the corresponding author.
    His email address is \href{mailto:e.costantini@tilburguniversity.edu}{e.costantini@tilburguniversity.edu}.
    The code used for the study is available on the author's GitHub page (https://github.com/EdoardoCostantini/mi-hd), or in more permanent form on Zenodo \citep{costantini:2023d}.
    Please read the README.md file for instructions on how to replicate the results.
    The EVS data used in this study are openly available in the GESIS Data Archive at https://doi.org/10.4232/1.13511 and should be downloaded independently.
    The article is also accompanied by an interactive results dashboard developed as an R Shiny app~\citep{costantini:2023c}.
    We encourage the interested reader to use this tool while reading the results and discussion sections.
    A user manual is included as a README file in the folder accessible through the DOI provided in the citation.
    The Shiny app can be downloaded, installed, and used as an R package.

\bibliography{./bib/bibshelf.bib}

\end{document}